\documentclass[twocolumn,twocolappendix]{aastex631}

\usepackage{newtxtext,newtxmath}
\usepackage[T1]{fontenc}
\usepackage{ulem}

\usepackage{graphicx}
\usepackage{amsmath}
\usepackage{xspace}
\usepackage{xcolor}
\usepackage{listings}
\usepackage{lipsum}
\usepackage{threeparttable}

\usepackage[all]{hypcap}

\newcommand\tabletophlineone{\hline\\[-0.16in]\hline\\[-0.15in]}
\newcommand\tabletophlinetwo{\hline\\[-0.15in]}

\newcommand\starsmasher{\texttt{StarSmasher}\xspace}

\makeatletter
\newcommand{\simulationlabel}[2]{\edef\@currentlabel{#1}\label{#2}}
\newcommand{\opacitylabel}[2]{\edef\@currentlabel{#1}\label{#2}}

\let\@makefntextorig\@makefntext
\renewcommand\@makefntext[1]{\hspace*{-3mm}\@makefntextorig{#1}}

\makeatother

\shorttitle{Obtaining a light curve}
\shortauthors{Hatfull \& Ivanova}

\begin{document}
\title{Simulating a stellar contact binary merger -- II. Obtaining a light curve\footnote{See the prior work in this series \citep{2021MNRAS.507..385H}.
}}

\author[0000-0003-0226-546X]{Roger W.M. Hatfull}
\affiliation{Department of Physics, University of Alberta, Edmonton, T6G 2E7, Alberta, Canada}

\author[0000-0001-6251-5315]{Natalia Ivanova}
\affiliation{Department of Physics, University of Alberta, Edmonton, T6G 2E7, Alberta, Canada}

\correspondingauthor{Natalia Ivanova}
\email{nata.ivanova@ualberta.ca}

\begin{abstract}
Luminous Red Novae (LRNe) are enigmatic transient events distinguished by a rapid rise in luminosity, a plateau in luminosity, and spectra which become redder with time. 
The best-observed system before, during, and after the outburst is V1309~Sco.
We model a candidate V1309~Sco progenitor binary configuration ($1.52+0.16\,M_\odot$) using the Smoothed Particle Hydrodynamics (SPH) code \starsmasher with a modified energy equation that implements flux-limited emission-diffusion radiative transport in a Lagrangian case.
We developed an imaging technique allowing us to capture the flux an observer would measure.
In this novel method, the outgoing radiative flux of each SPH particle in the observer's direction is attenuated by other particles along the path to the observer.
We investigated how the light curve is affected in various models: with and without dust formation; constant, Planck, or Rosseland mean opacities; different donor star sizes; different companion star masses and types; radiative heating included in our modified energy equation; and different SPH simulation resolutions.
The resulting evolution in bolometric luminosity and spectrum peak temperature is in good agreement with V1309~Sco observations.
Our simulations rule out V1309~Sco models that do not assume dust formation. 
\end{abstract}


\section{Introduction}

Luminous Red Novae (LRNe) are transient events discovered at the end of 20th century, with numerous events identified to date  \citep[e.g.][and more]{1990ApJ...353L..35M,1999AJ....118.1034M,2003Natur.422..405B, 2007Natur.447..458K, 2007Natur.449E...1P, 2009ApJ...695L.154B, 2011ApJ...737...17B, 2011ApJ...730..134K, 2011A&A...528A.114T,2015A&A...578L..10K,2017ApJ...834..107B,2019A&A...632L...6C,2019A&A...630A..75P,2021A&A...653A.134B,2021A&A...646A.119P,2021A&A...647A..93P,2022Univ....8..493C,2022A&A...667A...4C}.
LRNe outbursts have no strict definition but can be described as having peak luminosities somewhere between a nova and a supernova \citep{2012PASA...29..482K}, a plateau in luminosities after the main peak, and spectra which evolve from visual or near-infrared to redder colors over time.
In some LRNe, the apparent photospheric expansion velocity, as determined by the evolution of luminosity and temperature, is a factor of a few less than that derived from spectral lines.
LRNe exhibit unique spectral features, with prominent emission lines of hydrogen and other elements, providing insights into the physical processes at play during these stellar events \citep{2015A&A...580A..34K}.
LRNe are potentially produced by common envelope (CE) events \citep{2013Sci...339..433I}.
The pivotal observation that allowed a link to be made between CE events and LRNe was the outburst of V1309~Sco, for which a binary orbital period decay was observed prior to its outburst \citep{2011A&A...528A.114T}.

CE evolution is a theoretically proposed rapid phase of binary evolution required to explain the transformation of originally wide binaries into the observed close binaries \citep{1976IAUS...73...75P}.
While initially conceived to explain cataclysmic variables, it is now used to explain the formation of various astrophysically significant binaries, including X-ray binaries \citep{1976IAUS...73...35V}, binary pulsars \citep{1976ApJ...207..574S}, close double white dwarf binaries, Type Ia supernovae progenitors \citep{1993PASP..105.1373I}, hot subdwarfs \citep{2002MNRAS.336..449H}, and progenitors of gravitational wave sources \citep{1979IAUS...83..401T}.

During a CE event, a companion star plunges into the envelope of the original donor star, thereby enveloping the system in a CE.
Drag forces within the CE dissipate the binary's orbital energy, rapidly reducing the separation between the companion star and the donor's core. 
The outcome of a CE event can result in the formation of a more compact binary or the merger of the two stars.
In the case of the V1309~Sco progenitor, a merger is anticipated because no post-outburst binary has been detected yet.
However, the progenitors of other observed LRNe remain uncertain.
Whether a CE event leads to binary formation or merger is a subject of ongoing research.
Regardless of the progenitor's fate, the binary is expected to eject some of the donor's material, potentially forming post-CE planetary nebulae.
During the rapid expansion and cooling of the ejecta, the LRNe outburst occurs. 
The outburst carries information about the ejected material's quantity, entropy, and velocity.
Understanding the physics of LRNe is crucial for unravelling the mysteries of CE events.

A primary challenge in bridging theory with observations of LRNe lies in generating synthetic light curves from numerical simulations.
Currently, two approaches are prevalent to model a CE event itself: three-dimensional (3D) hydrodynamical simulations and simplified pseudo-1D simulations.
While 3D simulations offer detailed, comprehensive models, pseudo-1D simulations rely on assumptions about the stratified gas temperature.
However, both methods encounter complexities in accurately replicating LRNe behavior.
For pseudo-1D codes, a key challenge is establishing initial and boundary conditions for the ejected material, as those are not explicitly modeled \textit{in situ}.
For example, \cite{2024ApJ...963L..35C} employed proper radiative transfer in a 1D framework for a spherically symmetric geometry, but they had to introduce the ejecta in a highly simplified manner.
Furthermore, no self-consistent 3D simulation of a CE event has produced a light curve during the event itself to date.

In this work, we introduce a flux-limited emission-diffusion (FLED) technique to \starsmasher, which is a 3D Smoothed Particle Hydrodynamics (SPH) code specifically suited for stellar and planetary mergers \citep{2010MNRAS.402..105G, 2018ascl.soft05010G}.
SPH is a Lagrangian method in which a continuum is discretized into mass elements \citep{1977MNRAS.181..375G}, and for which, notably, radiative heat diffusion was an early consideration \citep{1977AJ.....82.1013L, 1985PASA....6..207B}.

Flux-limited diffusion (FLD) radiative transfer, first introduced to astrophysics by \cite{1981ApJ...248..321L}, is commonly implemented using the approach outlined in \cite{2007ApJ...667..626K}. There are implementations of FLD in SPH codes, often incorporating SPH summation techniques \citep[e.g., see][]{ 2004MNRAS.353.1078W, 2005MNRAS.364.1367W}.
The FLD method is well suited for simulating optically thick fluids such as those found in supernova explosions \citep[e.g.][]{1994ApJ...435..339H,2006ApJ...643..292F}, collapsing protostars \citep[e.g.][]{2006ApJ...639..559V}, protoplanetary disks \citep[e.g.][]{2007ApJ...661L..77M}, and more.
However, SPH summation cannot provide the outgoing radiative flux from the surface, and alternative models have been proposed in which particles are matched to profiles of objects for which the surface properties are known ahead of time, such as polytropic pseudo-clouds \citep{2009MNRAS.394..882F, 2015MNRAS.447...25L} or \texttt{MESA} stars \citep{2021MNRAS.507..385H}.
Escaping radiation from an SPH particle could deposit its energy into other particles long distances away and must be processed using other methods, such as ray tracing \citep{2000MNRAS.315..713K} or Monte Carlo \citep{2003MNRAS.343..900O}, though these tend to be computationally expensive and difficult to implement.

Our FLED technique, which we implement in \starsmasher, operates on similar length scales as SPH summation, and models the outgoing flux from each individual SPH particle based on the particle's local neighborhood, which is angle-dependent. This method acts as an intermediary between the approach used in \cite{2022ApJ...938....5M}, which focuses on diffusion processes between neighboring pseudo-1D shells, and FLD technique, as outlined in \cite[e.g., see][]{2007ApJ...667..626K}. We advance the approach in \cite{2022ApJ...938....5M} by implementing radiative energy transport directly between neighboring SPH particles, rather than pseudo-1D shells, and by incorporating radiation transport in optically thin regions, similar to FLD codes. Compared to more general FLD methods, our approach assumes a static regime (where gas velocities are much smaller than the speed of light), and neglects radiation feedback on the momentum equation. A detailed comparison of our FLED method and FLD method is provided in \S\ref{sec:meth_comp}.
We describe a novel post processing imaging technique to capture the propagation of radiative flux over long distances toward an observer, which we use to derive synthetic light curves and other observables such as the time evolution of velocity and spectra.

It has been argued that the light curves of LRNe are influenced by recombination processes \citep{2013Sci...339..433I}.
In our SPH code, we use a tabulated equation of state derived from {\tt MESA} \citep{2011ApJS..192....3P,2013ApJS..208....4P, 2015ApJS..220...15P, 2018ApJS..234...34P, 2019ApJS..243...10P, 2023ApJS..265...15J}, facilitating the accurate representation of matter capable of undergoing recombination of helium and hydrogen. 
Applying our radiative transfer method, we investigate the merger of a binary system that is often proposed as the progenitor of V1309~Sco.
Our analysis probes the influence of varying physics within our implementation on the resulting outcomes, particularly focusing on the light curve.
Finally, we compare our findings with the observed V1309~Sco outburst to validate our model. 

Specifically, in \S\ref{sec:radiativecooling}, we present our model of radiative cooling, encompassing both optically thick and thin cases.
Then, in \S\ref{sec:sph}, we delve into the specifics of implementing this model within an SPH code, and how we obtain observables, such as synthetic light curves.
\S\ref{sec:methodsverification} is dedicated to assessing the performance of our method, evaluating its ability to resolve the photosphere, its behavior in scenarios involving both stationary stars and dynamically ejected particles, and comparison to FLD.
Our SPH simulations are introduced in \S\ref{sec:simulations}, followed by a description of the main stages of our analysis in \S\ref{sec:results_stages}. 
We explore the impact of adopted opacities in \S\ref{sec:roleofopacities} and other model adjustments in \S\ref{sec:roleofvariedphysics}.
In \S\ref{sec:comparisonwithv1309sco} we present a comparative analysis with V1309~Sco and in \S\ref{sec:conclusions} we summarize our work.

\section{Radiative cooling}\label{sec:radiativecooling}

We start with the assumption that each SPH particle can be seen as an individual isothermal gas cloud (the SPH realization of this assumption will be discussed in \S\ref{sec:sph_realiz}).
More specifically, it is not the uniform temperature that is important in the considerations of radiative cooling in this section but rather the uniformity of the radiation energy density throughout the particle.
While each particle is itself isothermal, different particles have different temperatures, forming a spatial temperature gradient.
We note that assigning a single temperature to an individual particle is similar to assigning a single temperature value to a mesh in mesh-based computational methods.

\subsection{Emerging radiation.}\label{sec:radwithLTE}

The formal solution for the radiative transfer equation, assuming wavelength-independent opacities, is:
\begin{multline}\label{eq:firstI}
    I(\tau,\mu) = I(\tau_c,\mu) e^{-(\tau_c-\tau)/\mu} \\
    + \frac{1}{\mu}\int^{\tau_c}_\tau S(\tau^\prime,\mu) e^{-(\tau^\prime - \tau)/\mu} d\tau^\prime.
\end{multline}
Here $\tau$ is the optical depth for which we find the intensity of radiation $I(\tau,\mu)$ in the direction $\mu=\cos\theta$, where $\theta$ is the colatitude angle, $\tau_c$ is the total optical depth of the cloud, and $I(\tau_c,\mu)$ is the intensity of the radiation that illuminates the cloud.
The emerging intensity is that which is emitted from the surface of the gas cloud at $\tau=0$:
\begin{multline}\label{eq:Isurfintermediate}
    I_\text{emerg}(\mu) = I(\tau_c,\mu)e^{-\tau_c/\mu} \\
    + \frac{1}{\mu}\int^{\tau_c}_0 S(\tau^\prime,\mu) e^{-\tau^\prime/\mu} d\tau^\prime\ .
\end{multline}
From the approximation of being isothermal, the cloud is in a local thermal equilibrium described by some single temperature value, and hence the source function $S(\tau,\mu)$ is the Planck blackbody function $B(T)=\sigma T^4/\pi$:
\begin{align}\label{eq:Isurf}
    I_\text{emerg}(\mu) &= I(\tau_c,\mu)e^{-\tau_c/\mu} + B(T) \left(1-e^{-\tau_c/\mu}\right)\ .
\end{align}

By definition, the intensity is related to the rate with which radiative energy $E_\text{rad}$ at the optical depth $\tau$ passes through an area $dA$ into a solid angle $d\Omega$:
\begin{align}\label{eq:intensity}
    I (\tau, \mu) \equiv \frac{dE_\text{rad}(\tau )}{dt} \frac{1}{\mu dA} \frac{1}{d\Omega}\ ,
\end{align}
where $d\Omega = -d\mu d\phi$, and $\phi$ is the polar angle.
Using Equation~(\ref{eq:intensity}), we integrate over outgoing directions ($\mu \ge 0$) and over the entire surface of the gas cloud to obtain the rate at which energy is radiated away from the surface of the cloud
\begin{align}
    \frac{dE_{\rm emerg}}{dt} &= 2\pi A \int_0^1 I_\text{emerg}(\mu)\mu\, d\mu\ ,\label{eq:dEdttot}
\end{align}
where $A$ is the surface area of the cloud and $0\leq\mu\leq1$ is the outward direction.
Combining Equations~(\ref{eq:dEdttot}) and~(\ref{eq:Isurf}) yields
\begin{eqnarray}
    \frac{dE_{\rm emerg}}{dt} = 2\pi A \left[ \int_0^1 I(\tau_c,\mu)e^{-\tau_c/\mu} \mu\, d\mu +  B(T) Q(\tau_c) \right]\ .\label{eq:dEdttotfinal}
\end{eqnarray}  
Here we introduced a new quantity:
\begin{eqnarray}
    Q(\tau_c) &\equiv& \int_0^1 \mu\left(1-e^{-\tau_c/\mu}\right)d\mu \nonumber \\
    &=& \frac{1}{2}\left[1 - (1-\tau_c)e^{-\tau_c} + \tau_c^2 \text{Ei}\left(-\tau_c\right)\right] \ ,\label{eq:Q}
\end{eqnarray}
where $\text{Ei}$ is the exponential integral function.
In optically thick and optically thin regimes, $Q(\tau_c)$ approaches asymptotic values: 
\begin{eqnarray}
    \lim_{\tau_c\ll1} Q(\tau_c) &=& \tau_c \ , \label{eq:Qthin} \\
    \lim_{\tau_c\gg1} Q(\tau_c) &=& \frac{1}{2} \nonumber .\label{eq:Qthick}
\end{eqnarray}
In our implementations discussed in future sections, we use pre-calculated values of $Q(\tau_c)$ if $0.01 \le \tau_c \le 5$, and the limiting values as above otherwise.

In the case of no incoming radiation, radiative energy losses are simply 
\begin{equation}
    \frac{dE_{\rm emerg}}{dt} = 2 A \sigma T^4  Q(\tau_c) = \frac{1}{2} A c U_{\rm rad}  Q(\tau_c)  \ ,
    \label{de_emerg_dt}
\end{equation}
where $U_{\rm rad}=aT^4$ is the radiative energy density.
In the optically thick regime, the emergent radiation becomes the standard blackbody radiation
\begin{equation}
    \lim_{\tau_c\gg1} \frac{dE_{\rm emerg}}{dt} = A \sigma T^4 \ , \label{eq:dEdtthick}
\end{equation}
Albeit equation~\ref{eq:dEdtthick} can be deemed accurate mathematically, from a physical perspective, the limit implied by this equation is not applicable, as the derivation assumes an optically thin environment. The optically thick scenario, where photons are trapped and must scatter before escaping, is discussed in detail in \S\ref{sec_dif}.
In the optically thin regime, the radiative cooling becomes 
\begin{equation}
    \lim_{\tau_c\ll1} \frac{dE_{\rm emerg}}{dt} = 2A \sigma T^4 \tau_c  \ .    \label{eq:dEdtthin}
\end{equation}

\subsection{Diffusion limited cooling}
\label{sec_dif}

In a group of optically thick objects, radiative energy transfer can only proceed in the presence of a nonzero temperature gradient.
Without the temperature gradient, the radiative flux is zero.
An individual object surrounded by other objects of the same temperature cannot cool by radiation and will remain in equilibrium.

The time it takes for a  photon to diffuse out of an object with the size $r$ by a random walk is 
\begin{equation}
    t_{\rm diff}= \frac{r^2}{D} =  \frac{\kappa \rho r^2 }{c} = \frac{\tau r}{c}\ ,
\end{equation}
where $D\equiv c l_{\rm ph} = c/\kappa \rho$ is the diffusion coefficient, $l_{\rm ph}$ is the photon's mean free path, $c$ is the speed of light, $\rho$ is the density, and $\kappa$ is the opacity.

The rate at which the object can radiate away energy is limited by
\begin{equation}
    \frac{dE_{\rm max, diff}}{dt} = \frac{E_{\rm int}}{t_{\rm diff}} =  \frac{m u}{t_{\rm diff}}\ ,
    \label{eu_diff}
\end{equation}
where $E_{\rm int}$ is the total internal energy of the object, $m$ is its mass, and $u$ is its specific internal energy.
In our case, the object is an SPH particle.

The actual energy loss rate will be smaller.
The luminosity is the flux through a given area $A$ due to photon diffusion in the presence of an existing gradient of radiative energy: 
\begin{equation}
    L_{\rm diff} =  - A \ \frac{D}{3} \ \frac{d (U_{\rm rad})}{dr}=  - A \ \frac{D}{3} \ \frac{d (u_{\rm rad} \rho)}{dr} \ ,
\end{equation}
where $u_{\rm rad}$ is specific radiative energy and $1/3$ is the geometrical factor common for diffusion theory.
This diffusion approximation equation can be recognized if rewritten in its more widely used form, known as the equation of radiative transport that operates in a spherically symmetric optically thick environment inside a star: 
\begin{equation}
    L_{\rm diff} = - 4\pi r^2  \frac{c}{3 \kappa \rho} \frac{d aT^4}{dr} \ .
\label{eq:ldiff_1d}    
\end{equation}

We hence introduce diffusion-limited radiative cooling as the rate at which the radiative energy departs due to the local gradient of radiative energy density:
\begin{equation}
    \frac{dE_{\rm diff}}{dt} =  - \frac{1}{3} A\ D\  \nabla \left (a T^4 \right )=  - \frac{1}{3} A\ D\   \nabla U_{\rm rad} \ .
    \label{edot_diff}
\end{equation}
The radiative energy loss from an SPH particle cannot exceed the value above.
Only a negative temperature gradient results in the radiative energy leaving the particle.
As long as the particle is optically thick with a photon mean free path much less than its size, then ${dE_{\rm diff}}/{dt} < {dE_{\rm emerg}}/{dt} $.

We introduce $\beta$ as the ratio of radiative specific energy and total specific internal energy:
\begin{align}
\beta &= \frac{u_{\rm rad}}{u} = \frac{a T^4}{\rho u}\ , & 0\le\ &\beta \le 1\ .
\end{align}
Note that this  $\beta$ is not the same $\beta_g$ often used to show the role of the gas pressure in the total pressure.
Equation~(\ref{edot_diff}) can be rewritten as:
\begin{equation}
    \frac{dE_{\rm diff}}{dt} =  - \frac{1}{3} A\ D\   \nabla \left (\beta u \rho \right )\ .
\end{equation}

The ability of the SPH particle to lose radiative energy into its surrounding depends on the local radiative energy density $\nabla U_\text{rad} = \nabla aT^4$.
We anticipate that this gradient depends on the direction in a 3D asymmetric gas cloud.

To obtain a quantification of cooling properties, we take samples of $\nabla U_\text{rad}$ in several directions.
We term each direction as a ``ray''.
We build $N_\text{ray}$ rays, each positioned at the center of the particle, and evenly spread in their directions.
To achieve an even distribution of ray directions, we sample the vertices of a regular icosahedron, subdivided $d$ times for $N_\text{rays}(d)=10(2^d)^2+2$.
We use $d=0$ for $N_\text{rays}=12$ in our simulations. 

We define $\vec{R}_{i,s}$ as the position at the distance $r_{\rm out}$ from particle $i$ in the direction of a ray with index $s=1,...\,,{N_{\rm rays}}$.
At that position $\vec{R}_{i,s}$, we find the radiation energy density  $U_{\rm rad}(\vec{R}_{i,s})$ produced by other particles (see more detail in \S\ref{sec:sph_realiz} on the choice of $r_{\rm out}$ and how $U_{\text{rad}}(\vec{R}_{i,s})$ is reconstructed).
In Figure~\ref{fig:raysketch} we show in red a sketch of an SPH particle with a ray $\vec{R}_{i,s}$ as a black arrow. 

If there is matter in the direction of the ray, and $U_{\text{rad}}(\vec{R}_{i,s})>0$, then the radiative energy density  gradient in that direction is found as
\begin{equation}\label{nabla_ur}
\nabla_s U_{\rm rad, i} = \frac{U_{\text{rad}}(\vec{R}_{i,s}) - U_{\rm rad, i}}{r_{\rm out}} \ .
\end{equation}
Only rays with negative temperature gradient contribute to the total radiative cooling of the particle $i$.
The total rate is then found as the sum of the contribution from each ray:
\begin{equation}
    \frac{dE_{\rm diff}}{dt} =  - \frac{1}{3} \frac{A}{N_\text{rays}}\ D\ \sum_{s=1}^{N_{\rm rays}} \nabla_s U_{\rm rad, i} \ ,
    \label{dediff_dt_nabla}
\end{equation}
where $A/N_\text{rays}$ is the total surface area through which each ray radiates.

\subsection{The final cooling rate}\label{sec:thefinalcoolingrate}

The total radiative energy loss is
\begin{equation}\label{eq:dEraddt}
    \frac{dE_{\rm rad}}{dt} =  \min\left ( \frac{dE_{\rm emerg}}{dt}, \frac{dE_{\rm diff}}{dt}, \frac{dE_{\rm max, diff}}{dt}\right ) \ .
\end{equation}
The radiative cooling term, or the rate at which the specific internal energy of an SPH particle changes with time, is 
\begin{equation}
    \dot u_{\rm rad, cool} = 
    \frac{du_{\rm rad}}{dt} = 
    - \frac{1}{m} \frac{dE_{\rm rad}}{dt}\ .
    \label{urad_dot}
\end{equation}
As we are finding cooling rates $\dot u_{\rm rad, cool}(t)$ for all particles at each moment of time $t$ separated by time step intervals $\Delta t$, we adopt that only the SPH particles with $\Delta t/t_{\rm diff}(t) \ge 10^{-12}$ can cool radiatively during our simulations.
If $\Delta t/t_{\rm diff}(t) < 10^{-12}$, we adopt that  $\dot u_{\rm rad, cool}(t)=0$.

It is important to note that the radiative energy loss calculated above represents the radiative energy emitted from the particle. As this energy leaves the particle, it can be absorbed by other particles along its path, facilitating radiative energy transport. In \S\ref{sec:heating}, we will explain how we account for the absorption of incoming radiative energy from surrounding particles, which contributes to the local heating of each particle.

\section{Radiative cooling and an SPH consideration}\label{sec:sph}

\begin{figure}
    \centering
    \includegraphics{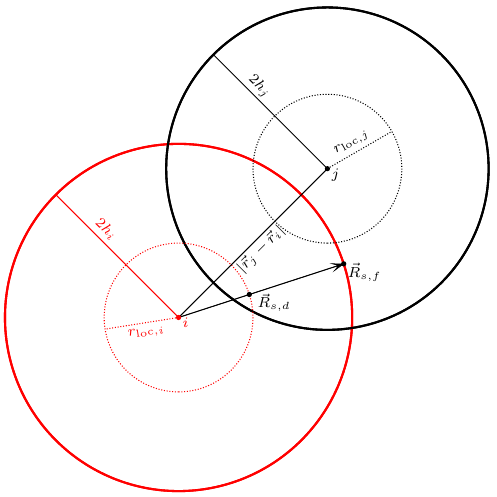}
    \label{fig:raysketch}
    \caption{A 2D sketch of an SPH particle $i$ (red) and another nearby particle $j$ (black) separated by distance $|\vec{r}_j-\vec{r}_i|$. We show a ray (black arrow) emerging from particle $i$ along direction $\vec{R}_s$. In the ``fluffy'' case (solid lines), position $\vec{R}_{s,f}$ is encapsulated by the kernel of particle $j$, which has size $2h_j$. However, in the ``dense'' case (dotted lines), the position $\vec{R}_{s,d}$ is not encapsulated, as the particle sizes are $r_{\text{loc},i} = (3m_i / 4\pi\rho)^{1/3}$. In the fluffy case, the temperature is nonzero at $\vec{R}_{s,f}$. In the dense case, the temperature is zero at $\vec{R}_{s,d}$.}
\end{figure}

\subsection{Finding quantities for SPH particles}\label{sec:sph_realiz}

The SPH method operates with so-termed primary variables and derivable quantities.
Specifically, the masses of the particles and their energies are the primary variables engaged in the conservation equations.
On the other hand, density, for example, is a derivable quantity: it is reconstructed using the masses of the particles in the neighborhood, their smoothing lengths, and the adopted kernel function.
An SPH particle $i$ located at $\vec{r}_i$ has a number of neighbors $N_{\text{nb},i}$, each of which satisfy $|\vec{r}_j - \vec{r}_i| < 2h_i$, where $h_i$ is the smoothing length and the kernel function $W$ has compact support $2h$.
Any derivable quantity $A_i$ can be found at location $\vec{r}_i$:
\begin{align}\label{eq:sphestimator}
    A_i &= \sum_j^{N_{\text{nb},i}} m_j \frac{A_j}{\rho_j} W(|\vec{r}_j-\vec{r}_i|, h_i) \ .
\end{align}
For example, the density at the position of particle $i$ is written as
\begin{align}
    \rho_i = \sum_j^{N_{\text{nb},i}} m_j W(|\vec{r}_j-\vec{r}_i|, h_i)\ .
\end{align}
However, the SPH summation method cannot guarantee a proper value of $\rho$ at any arbitrary location.
Unfortunately, this introduces unavoidable ambiguity in how the radiative cooling terms can be obtained, and below we outline two approaches.

In both approaches, we use SPH quantities $\rho_i$ and $u_i$ to find $T_i$ at the location of the particle $i$, using our tabulated equation of state, and then we find $\beta_i$.
Our primary concern is that an SPH particle should not lose more radiative energy than it possesses or more than it can lose given by the background conditions.
Hence, the difference in the two methods we will consider is in what is adopted as the average radiation field through particle $i$ and what is the gradient of radiation density in a given direction.

We will consider two limiting cases.
In one, the radiation field is uniform within a particle's kernel, which in our case has radius $2h_i$ (due to our kernel function having compact support $2h$).
We will call this approach ``fluffy''.
In the other method, which we call ``dense'', the size of the uniform radiative field is determined by the central radiative energy density and the total energy that the particle has.
We will then compare the outcomes.
In the illustration shown in Figure~\ref{fig:raysketch}, the fluffy approach is drawn with solid lines, and the dense approach with dotted lines.

In what follows, we will use subscripts $i$ and $j$ to indicate quantities obtained using the SPH summations method for $i$ or $j$ SPH particles, those are not dependent on the adopted approach.
Subscripts $f$ or $d$ are used to indicate quantities dependent on the adopted approach, fluffy or dense, accordingly.

\subsubsection{Fluffy}

The particle size $r_{\rm out}=2h_i$ in this case.
The total internal energy per particle is $m_iu_i$.
The volume-averaged radiative energy density that influences the cooling rate in this approach is then 
\begin{equation}
    \overline{U}_{\rm rad,f} = \overline{(aT_i^4)} = \frac{3\beta_i u_i m_i }{4 \pi (2 h_i)^3}\ .
\end{equation}
This is the quantity we use for $U_{\rm rad,i}$ in Equations~(\ref{de_emerg_dt}) and~(\ref{nabla_ur}). It can be seen that $\overline{U}_{\rm rad,f} = \beta_i u_i \overline{\rho}_i$, where $\overline{\rho}_i =  3m_i/4 \pi (2 h_i)^3$.

We consider here a sphere that is uniformly lit with radiation.
A photon trying to escape the sphere undergoes a random walk process under the conditions imposed by the local density $\rho_i$, temperature $T_i$, and opacity $\kappa_i$.
All the related quantities, including the emerging radiation, can be found:
\begin{eqnarray}\label{eq:fluffystuff}
    A_f &=& 4 \pi ( 2h_i)^2\ , \\ 
    D_f &=& \frac{c}{\kappa_i \rho_i}\ , \nonumber  \\
    t_{\rm diff,f} &=& \frac{\kappa_i \rho_i (2 h_i)^2}{c}\ , \nonumber  \\
    \tau_f &=& \rho_i \kappa_i 2 h_i\ , \nonumber  \\
    \frac{dE_{\rm emerg,f}}{dt} &=& \frac{1}{2} A_f c \overline{U}_{\rm rad,f}  Q(\tau_f)\ , \nonumber \\
    \frac{dE_{\rm max, diff,f}}{dt} &=& \frac{u_i m_i }{t_{\rm diff,f}}\ . \nonumber 
\end{eqnarray}

The diffusion limited cooling rate is more complicated here as the radiation energy density that is associated with the particle $i$ does not represent the true (summed from many particles) radiation energy density at this location, and hence cannot be used directly to find $\nabla U_{\rm rad}$.

The radiation energy density is a steep function of temperature: when the temperature drops by only half, the radiation energy density drops by 16 times.
A typical ratio between particle size and distance to the binary $d$ in our simulation is about $0.1-0.2$.
One cannot expect a monotonically dropping temperature profile in the ejecta, as some particles could be shocked more than others.
Nonetheless, a change by a factor of two or more in temperature between the particles at a distance of $\sim 0.1d-0.2 d$ can be expected. 

The steps for obtaining the cooling rate are as follows:
\renewcommand{\theenumi}{(f.\roman{enumi})}
\begin{enumerate}
\item For each ray at particle $i$, we find particle $j\neq i$ closest to location $\vec{R}_{i,s}$ and whose kernel encapsulates $\vec{R}_{i,s}$.

\item \label{fluffytemperaturecheck1} If temperature $T_j<T_i$, or there is no such particle $j$, then
\begin{equation}\label{eq:gradUradfluffynonearby}
\nabla_s U_{\rm rad,f} = - \frac{\overline{U}_{\rm rad,f}}{2h_i}\ .
\end{equation}

\item \label{fluffytemperaturecheck2} If $T_j\geq T_i$, no diffusion-limited cooling takes place in that direction,
$ \nabla_s U_{\rm rad,i}=0$.

\item The diffusion limited cooling rate is found as 
\begin{equation}
    \frac{dE_{\rm diff,f}}{dt} =  \beta_i \frac{dE_{\rm max, diff,f}}{dt} \times \frac{1}{ N_{\rm rays}}  \sum_{s=1}^{N_{\rm rays}}  C_s\ ,
\end{equation}
where $C_s=1$ if diffusion-limited cooling takes place in the direction of ray $s$ and $0$ otherwise.
\end{enumerate}
\renewcommand{\theenumi}{(\roman{enumi})}

Physically, the fluffy approximation implies that each SPH particle will cool at a rate corresponding to the uniform radiation energy it has inside its radius $2h_i$, so the number of photons that can start leaving the particle depends on the particle's own radiation energy density.
On the other hand, the photon departure rate is influenced by the density provided by all the particles and overall opacity that photon experiences -- leaving photons will scatter and diffuse through the media that is determined not just by particles' own matter.
Since the particles overlap, the resulting cooling rate at each point of space is the combination of several particles cooling simultaneously.

\subsubsection{Dense}\label{sec:dense}

At the location of a particle, the radiation energy density is exactly  
\begin{equation}
    U_{\rm rad,i}=aT_i^4 \ .
\end{equation} 
While in our fluffy approximation the volume-averaged radiation energy density is less than this value for an individual particle, the overlapping of particles in space brings the combined value up to the expected.
We can, however, investigate the case when each particle is an individual non-overlapping particle that has the same energy density throughout as in its center for the purpose of radiative losses.
To conserve the total energy per particle, its volume has to be adjusted.
The effective size of each particle is smaller than $2h_i$:
\begin{equation}\label{eq:rloc}
    r_{\rm loc} = \left ( \frac{m_i}{\frac{4}{3} \pi \rho_i } \right )^{1/3}\ .
\end{equation}
With this effective size, the volume-averaged radiative energy density that influences the cooling rate of particle $i$ is
\begin{equation}
U_{\rm rad,d}=U_{{\rm rad,}i}=aT_i^4 \ . 
\end{equation}

The sampled locations $\vec{R}_{i,s}$, are in this case located at distance $r_{\rm out}=r_{\rm loc}$.
The value of $U_{\text{rad}}(\vec{R}_{i,s})$ at these locations is found in the following manner:
\renewcommand{\theenumi}{(d.\roman{enumi})}
\begin{enumerate}
    \item For each ray at particle $i$, we find particle $j\neq i$ closest to $\vec{R}_{i,s}$ with $|\vec{r}_j - \vec{R}_{i,s}| < r_{\text{loc},j}$.
    \item \label{densetemperaturecheck1} If temperature $T_j < T_i$ then
    \begin{align}
        \nabla_s U_\text{rad,d} = \frac{U_{\text{rad},j} - U_{\text{rad},i}}{|\vec{r}_j - \vec{r}_i|}\ ,
    \end{align}
    or, if there is no such particle $j$, then
    \begin{align}\label{eq:gradUraddensenonearby}
        \nabla_s U_\text{rad,d} = -\frac{U_{\text{rad},i}}{r_{\text{loc},i}}\ .
    \end{align}

    \item \label{densetemperaturecheck2} If $T_j\geq T_i$, no diffusion-limited cooling takes place in that direction, $\nabla_s U_{\text{rad},i} = 0$.
    
    \item The diffusion limited cooling rate is found as
    \begin{align}
        \frac{dE_\text{diff,d}}{dt} &= -\frac{1}{3}\frac{A_d}{N_\text{rays}} D_d \sum_{s=1}^{N_\text{rays}} \nabla_s U_\text{rad,d}\ .
    \end{align}
\end{enumerate}
\renewcommand{\theenumi}{(\roman{enumi})}

The other necessary quantities are found:
\begin{eqnarray}
    A_d &=& 4 \pi r_{\rm loc}^2\ , \\ 
    D_d &=& \frac{c}{\kappa_i \rho_i}\ , \nonumber  \\
    t_{\rm diff,d} &=& \frac{\kappa_i \rho_i r_{\rm loc}^2}{c}\ , \nonumber  \\
    \tau_d &=& \rho_i \kappa_i r_{\rm loc}\ , \nonumber  \\
    \frac{dE_{\rm emerg,d}}{dt} &=& \frac{1}{2} A_d c {U}_{\rm rad,d} Q(\tau_d)  \ ,\nonumber \\
    \frac{dE_{\rm max, diff,d}}{dt} &=&   \frac{u_i m_i }{t_{\rm diff,d}}\ . \nonumber 
\end{eqnarray}

Suppose one splits a large object into several smaller-sized objects with the same total volume, density, and opacity.
In that case, the summative diffusion time from the several smaller objects will be shorter than the diffusion time from the single large object.
Hence, this approach will likely yield a cooling rate larger than both the fluffy method and reality.
The ratio between the dense diffusion-limited cooling rate and the fluffy diffusion-limited cooling rate for a particle that emits in empty space is on the order of $2h_i/r_{\rm loc} > 1$, which is about from $5$ to $10$ for most particles.
However, in the optically thin regime the emerging radiation from a particle is the same between the two approaches:
\begin{equation}
\frac{dE_{\rm emerg,d}}{dt} =\frac{dE_{\rm emerg,f}}{dt} =\frac{3}{2} c \beta_i u_i m_i \rho_i \kappa_i \ . 
\end{equation}

Additionally, we expect the total surface area exposed to empty space in the dense case to be larger on average than in the fluffy case, thereby increasing the flux.
We will therefore consider the cooling rate from the dense approach as the maximum possible and from the fluffy approach as the minimum possible.

\subsection{Heating}\label{sec:heating}
Previously, we considered the loss of energy by SPH particles from radiation.
Here we consider the absorption and reprocessing of the emitted radiation in the SPH particles.
For each SPH particle $i$ we split the total specific energy rate of change due to radiation into two components:
\begin{align}
    \dot{u}_{\text{rad},i} = \dot{u}_{\text{rad,cool},i} + \dot{u}_{\text{rad,heat},i}\ ,
\end{align}
where $\dot{u}_{\text{rad,cool},i}$ is as described in Equation~(\ref{urad_dot}).

We adopt the following assumptions on radiative heating:

\begin{enumerate}
\item Only diffusion-limited radiation flux is used to heat the nearby particles. 
The emerging radiative flux is considered to leave the simulation domain without being re-captured.

\item Each ray in our diffusion-limited cooling deposits its energy into the nearest particle $j$ with $T_j<T_i$, determined by its proximity to $\vec{R}_{i,s}$.

\item Particle $j$ can only absorb $1-e^{-\tau_j}$ of the radiative flux that passes through it.

\item For particle $j$, the total radiative heating rate $\dot{u}_{\text{rad,heat},j}$ is the sum of all the radiation energy sent to it by other particles.
\end{enumerate}

In the fluffy case
\begin{align}\label{eq:udotradheatfj}
    \dot{u}_{\text{rad,heat,f},j} = \frac{1 - e^{-\tau_j}}{m_j N_\text{rays}} \sum_i \beta_i\frac{dE_{\text{max,diff},i}}{dt}\ ,
\end{align}
and in the dense case
\begin{align}\label{eq:udotradheatdj}
    \dot{u}_{\text{rad,heat,d},j} = -\frac{1-e^{-\tau_j}}{3 m_j N_\text{rays}} \sum_i A_{\text{d},i} D_{\text{d},i} \frac{U_{\text{rad},j} - U_{\text{rad},i}}{|\vec{r}_j - \vec{r}_i|}\ .
\end{align}

\subsection{Opacities}\label{sec:opacities}

\begin{figure*}
    \centering
    \includegraphics{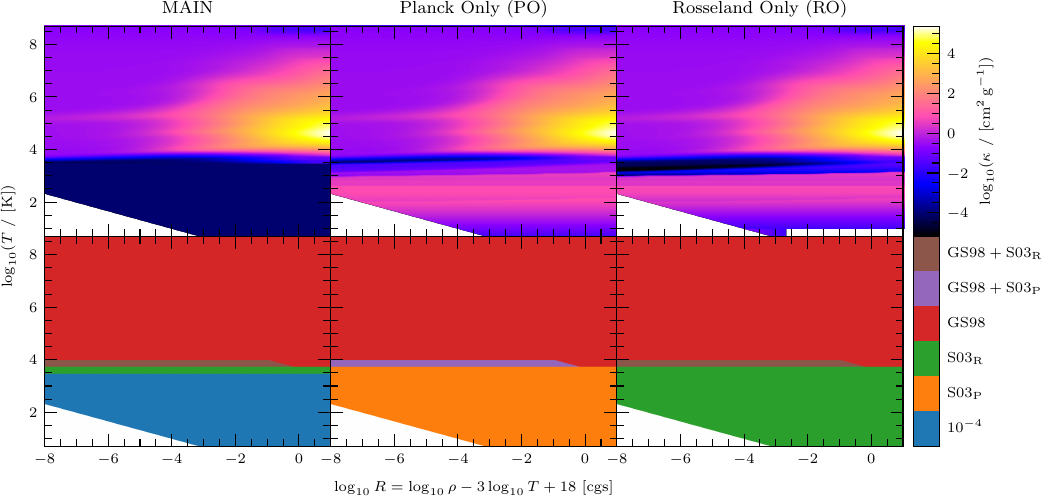}
    \label{fig:opacities}
    \caption{The three opacity tables (top panels) used in our dynamical simulations are shown in $\log T$, $\log R$ space. We show how each of our opacity tables are constructed following \S\ref{sec:opacities} (bottom panel): ``$10^{-4}$'' is the manually set $10^{-4}\,\text{cm}^2\,\text{g}^{-1}$ region (blue); $\text{S03}_\text{P}$ are Plank opacities from S03 for $T < 3\,000\,\text{K}$ (orange); $\text{S03}_\text{R}$ are Rosseland opacities from S03 for $T < 3\,000\,\text{K}$ (green); and GS98 (red).
    We also show the $3.75 < \log T<4$ blending regions between GS98 and $\text{S03}_\text{P}$ (purple) and between GS98 and $\text{S03}_\text{R}$ (brown).}
\end{figure*}

We create opacity lookup tables using a combination of three tables, which we call our ``source opacity tables'', and each of which have solar abundances $X=0.7$ and $Z=0.02$: \citet{1998SSRv...85..161G} (``GS98'' hereafter), as implemented in \texttt{MESA} version r10398 with file name ``\texttt{gs98\_z0.02\_x0.7.data}''); and \citet{2003A&A...410..611S} (``S03'' hereafter).
The source opacity tables cover the following domains:
\begin{align}
    \text{GS98} &\left\lbrace\begin{aligned}
        3.75 < &\log T < 8.7\\
        -8 < &\log R < 1
    \end{aligned}\right. & \text{S03}\left\lbrace\begin{aligned}
        0.7 < &\log T < 4\\
        -17.7 < &\log\rho < -6.7
    \end{aligned}\right.\ , \nonumber
\end{align}
where $\log R = \log\rho - 3\log T + 18$.
We note that S03 consists of two distinct source opacity tables: Planck mean opacities (``$\text{S03}_\text{P}$'' hereafter) and Rosseland mean opacities (``$\text{S03}_\text{R}$'' hereafter).

Each of our simulations uses a single opacity table.
The opacities of individual SPH particles are obtained using their densities and temperatures to perform a linear interpolation on the table in log space.
We describe our final three lookup tables that are used in our simulations and are shown in Figure~\ref{fig:opacities}.
We hereafter refer to the process of combining source opacity tables by linear interpolation in log space as ``blending.''
\begin{enumerate}%
\renewcommand{\labelenumi}{(\theenumi)}%
\renewcommand{\theenumi}{\arabic{enumi}}%
    \item\opacitylabel{MAIN}{fullopacities} {\bf MAIN:} Where $\log T \le 3.48$ ($T\le 3\,000\,\text{K}$), we set opacities to a constant low value of $10^{-4}\,\text{cm}^2\,\text{g}^{-1}$.
    We use $\text{S03}_\text{R}$ exclusively where $3.48 < \log T \le 3.75$, GS98 exclusively where $\log T \ge 4$, and a blend of $\text{S03}_\text{R}$ and GS98 where $3.75 < \log T < 4$.
    
    \item\opacitylabel{PO}{planckonlyopacities} {\bf PO:} We use $\text{S03}_\text{P}$ exclusively where $\log T \le 3.75$, GS98 exclusively where $\log T \ge 4$, and a blend of $\text{S03}_\text{P}$ and GS98 where $3.75 < \log T < 4$.

    \item\opacitylabel{RO}{rosselandonlyopacities} {\bf RO:} We use $\text{S03}_\text{R}$ exclusively where $\log T \le 3.75$, GS98 exclusively where $\log T \ge 4$, and a blend of $\text{S03}_\text{R}$ and GS98 where $3.75 < \log T < 4$.
\end{enumerate}

If radiation energy density is represented by the Planck function, then for radiative energy transfer, Rosseland mean opacities should be used across all temperatures \citep{1998cmaf.conf..161M}; this is our {\bf RO} case.
However, it is not yet fully established whether or not dust formed rapidly during the merger.
To explore this uncertainty, we created additional opacity tables with modified opacities for temperatures $T \le 3\,000\,\text{K}$.
Our table {\bf MAIN} represents the scenario where dust did not form.
Our table {\bf PO} represents the scenario where dust did form, but remained optically thin \citep[e.g., see][where Plank opacities were used for dust nucleation during dynamic dust formation]{2022AA...667A..75S}.
We use Planck mean opacities where $T \le 3\,000\,\text{K}$ in our table {\bf PO}.
In this study, we do not separate the transport and emission regimes, and so we apply our opacity tables to both.

\subsection{Imaging the flux}\label{sec:imaging}

We generate post-processed images of the flux emitted from our simulations for comparison to the V1309~Sco outburst.
We instantiate an observer in the simulation, positioned infinitely far away and looking along a specified viewing angle.
We build a collection of parallel rays on a 2D grid (one ray per cell) at the observer's position, pointing along the viewing angle (analogous to an orthographic projection).
Hereafter we use the reference frame of the observer when referring to physical positions, defining the so-called ``screen space'', in which the viewing angle is normal to the $xy$ plane and the $-\hat{z}$ direction points away from the observer, along their view (``into the screen'').
Each of the rays on the 2D grid are parallel to the $z$ axis and point in the $-\hat{z}$ direction.

For a given cell on the 2D grid, we label SPH particles as ``interacting'' if the cell ray passes through their kernel.
For each cell ray we obtain the flux emitted along that ray toward the observer as the sum of the contributions from all interacting particles $i$:
\begin{align}\label{eq:Fcell}
    F_\text{cell} &= \sum_i F_{i,\text{rad}} e^{-\tau_{i,\text{cell}}},
\end{align}
where $\tau_{i,\text{cell}}$ is the optical depth which the radiative flux $F_{i,\text{rad}}$ must pass through to reach the observer.
To obtain $\tau_{i,\text{cell}}$ we let $\kappa_i\rho_i$ be constant throughout each particle along the cell ray:
\begin{equation}
    \tau_{i,\text{cell}} = \sum_j \tau_j \ .
\end{equation}
We selectively ignore $F_{i,\text{rad}}$ contributions from particles for which $\tau_{i,\text{cell}} \ge 20$.

In the fluffy cooling regime, many particles whose kernels overlap in space create a cumulative cooling rate which approaches that in the dense regime.
However, as particles never share the same positions and sizes, a cell ray will first encounter a contribution of only one of the fluffy particles in the group before it encounters the rest.
When that first particle is optically thick, the cell ray will not provide the correct cumulative cooling rate at the ``surface''.
We therefore adopt that during image construction each particle has a size $r_\text{loc}$ (Equation~\ref{eq:rloc}) and a uniform radiation field defined by its location, regardless of whether the simulation was performed in the fluffy or dense regime.
We note that in most cases during the pre-plunge stage, the first particle along a cell ray is optically thick and thus determines the surface.
In such cases, that optically thick particle is the final particle that contributes to the cell ray.

The contribution to the total flux along the cell ray from a particle $i$ is
\begin{equation}\label{eq:Firad}
    F_{i,\text{rad}} = \frac{1}{A_d} \min\left(\frac{dE_\text{diff,d}}{dt},\ \frac{dE_\text{emerg,d}}{dt}\right)\ .
\end{equation}

We further take into account the angular distribution of the rays: not all directions result in successful cooling, such that the visible surface of a particle may produce only a fraction of the flux compared to $F_{i,\text{rad}}$.

There could be optically thin particles along the cell ray between the observer and an optically thick particle.
The essence of the adopted approximation implies that we attenuate the illumination of the optically thick particle by the optically thin particles along the cell ray, as the radiation field of the optically thin particles does not affect the gradient inside the optically thick particle.

The 2D grid is constructed with resolution $(N_x,\, N_y)$ on the $xy$ plane in screen space and usually fits the visible boundary of all particle kernels, though with special considerations detailed below.
For each grid cell we cast a ray in the $-\hat{z}$ direction and obtain the cell flux $F_\text{cell}$.
The total observed luminosity for comparisons to real observations is:
\begin{align}\label{eq:Lobs}
    L_\text{obs} &= 4\Delta x \Delta y\sum_\text{cell} F_\text{cell}\ ,
\end{align}
where $\Delta x$ and $\Delta y$ are the width and height of each grid cell, and the leading factor of $4$ is from the observer's assumption that the flux seen from one viewing angle is the same as what would be seen from any other viewing angle.

Capturing all the particles in our grid requires the grid domain to expand as time evolves.
The evolved outermost ejecta produces the least flux.
For the light curve calculations, we limit the grid domain to $2\,000\,R_\odot$ in the $xy$ plane, and we dynamically adjust the grid resolution with the evolution of the ejecta: we use $(2\,000,\,2\,000)$ for the first $18.5\,\text{days}$, $(1\,200,\,1\,200)$ for another $18.5\,\text{days}$ after, and $(800,\,800)$ until the end of the evolution.

To obtain a characteristic value of some quantity $H$ over the entire image, we use a flux-weighted average across all cell rays:
\begin{align}\label{eq:fluxweightedaverage}
    \langle H_\text{rad}\rangle = \frac{1}{F} \sum_\text{cell} H_\text{cell} F_\text{cell}\ ,
\end{align}
where the total flux $F\equiv \sum_\text{cell}F_\text{cell}$ and $H_\text{cell}$ is the value of $H_\text{rad}$ for any given cell in the image.
Here $\langle H_\text{rad}\rangle$ is the value of $H_\text{rad}$ which the infinitely far away observer would measure.
When $H_\text{cell}$ describes a quantity related to the SPH particles,
\begin{equation}\label{eq:fluxweightedaveragecell}
    H_\text{cell} \equiv \frac{1}{F_\text{cell}} \sum_i H_iF_{i,\text{rad}}e^{-\tau_{i,\text{cell}}}\ .
\end{equation}
For example, the expansion velocity
\begin{equation}
    \langle v_\text{exp} \rangle = \frac{1}{F} \sum_\text{cell} \sum_i |v_{z,i}| F_{i,\text{rad}} e^{-\tau_{i,\text{cell}}}\ ,\label{eq:averagevexp}
\end{equation}
where $v_{z,i}$ is the velocity of particle $i$ in the direction toward the observer.

To obtain the characteristic temperature of an extended object, we integrate the Planck function.
Specifically, a particle that is not obscured by any other particles on the line of sight to the observer has an effective temperature
\begin{equation}    
    T_{i,\text{eff}} \equiv \left(\frac{F_{i,\text{rad}}}{\sigma}\right)^{1/4}\ .
\end{equation}
For each particle separately, we adopt that its radiation is that of a blackbody and hence 
\begin{equation}
  \frac{\sigma}{\pi} T_{i,\text{eff}}^4 =  \int_0^{\infty} B_\lambda (T_{i,\text{eff}})d\lambda\ .
\end{equation}
We introduce the Planck weighting function $f_\lambda(T_{i,\text{eff}})$ as 
\begin{equation}
    f_\lambda(T_{i,\text{eff}}) \equiv \frac{\pi}{\sigma  T_{i,\text{eff}}^4} B_\lambda (T_{i,\text{eff}})\ .
\end{equation}
For this weighting function, $\int_0^\infty f_\lambda(T_{i,\text{eff}}) d\lambda=1$.

The spectral distribution of the apparent flux along a ray is the sum of the spectral contributions from each particle along that ray, 
\begin{equation}
    F_\lambda = \sum_{\rm cell}\sum_{i} f_\lambda(T_{i,\text{eff}}) F_{i,\text{rad}} e^{-\tau_{i,\text{cell}}}
\end{equation}
We term the spectrum peak temperature $T_\text{sp}$ as the temperature of a blackbody, where the spectral distribution of that blackbody peaks at the same wavelength as our resulting spectral distribution.
In this work, the spectrum peak temperatures are determined based on the peak of the spectral distribution $F_{\lambda,{\rm max}}$. The range of these peak temperatures is defined by the region where the spectral intensity remains within $95\,\text{per}\,\text{cent}$ of the peak value, specifically where $F_{\lambda} \ge 0.95F_{\lambda,{\rm max}}$.
Note that the resulting spectral distribution is not that of a blackbody (more about this in \S\ref{sec:results_stages}).

To reconstruct the spectral distribution we consider only $10\,\text{nm} < \lambda < 3\,000\,\text{nm}$ in steps of $10\,\text{nm}$, which includes the $V$-band, $R$-band, and $I$-band.
Our method can identify the peak and recover spectrum peak temperatures from about $1\,000\,\text{K}$ to about $100\,000\,\text{K}$.
We can also integrate the luminosity produced in specific bands using the spectral flux distribution.
We define the visual luminosity:
\begin{equation}\label{eq:Lobs_spectra}
    L_{\text{obs}, V} = 4 \Delta x \Delta y \int_{500 {\rm nm}}^{600{\rm nm}} F_\lambda d \lambda \ .
\end{equation}
$R$-band luminosity is found similarly in the ranges of $590-730\,\text{nm}$, and $I$-band luminosity is in the ranges of $720-880\,\text{nm}$\footnote{There is no single, uniformly used, wavelength ranges for $V, R$ and $I$ filters. We used values that are closely representing $V$ and $I$ filters used by OGLE, accordingly to \href{http://svo2.cab.inta-csic.es/theory/fps/}{svo2.cab.inta-csic.es/theory/fps/}}.

\section{Method's  verification and comparison}\label{sec:methodsverification}

\begin{figure}
    \centering
    \includegraphics{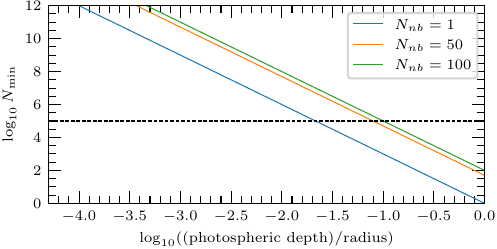}
    \label{fig:photosphericdepth}
    \caption{The minimum number of SPH particles $N_\text{min}$ in a close-packed hexagonal lattice required to resolve the photosphere of a star with some radius, given the photospheric depth (equation 34 of \citealt{2021MNRAS.507..385H}). All our simulations have $10 \lesssim N_\text{nb} \lesssim 100$ and $\sim10^5$ particles (dashed black line), and thus we expect to be able to resolve the photosphere once the depth is larger than $\sim0.1$ times the radius.}
\end{figure}

\subsection{A photosphere and its resolution}\label{sec:photosphere_resolution}

The term ``photosphere'' is somewhat vague and generally refers to the region from which the observed radiation is emitted. In our simulations and image processing, we do not employ a specific definition or predetermined location for the photosphere.  However, we can analyze how the final resolution affects what is typically considered the photosphere for a star.
We will also discuss why spatial resolution is less critical for particles where energy is carried by radiation, not convection.

For stars in hydrostatic equilibrium, the observed radiation is the same as if it had radiated from the optical depth of $\tau=2/3$.
For the case when particles occupy equal volume, one can find the minimum number of SPH particles required in order to represent a star with a resolved photosphere, in the sense that the outermost particles would be located at $\tau=2/3$ \citep[see][]{2021MNRAS.507..385H}. 
We show the relation between the minimum number of particles and the relative depth of the photosphere in Figure~\ref{fig:photosphericdepth}. 
The 1D \texttt{MESA} model that we use for our standard runs, $R_\text{1D}=3.715R_\odot$, has a photospheric depth of $\approx10^{-3}\,R_\odot$.
This implies that $N_\text{min}\gtrsim5\times10^{10}$ SPH particles are required to resolve the photosphere in the initial 3D stellar model.

However, the spatial resolution of the hypothetical photosphere, defined as some location where $\tau\le 1$, is crucial only if the photosphere's boundary conditions directly influence the outgoing flux. 
Analysis of energy transport equations shows that the role of photospheric boundary conditions varies between convection-dominated and radiation-dominated flux regimes. 
In radiation-dominated cases, the ``effective'' temperature is primarily determined by the radiative flux deep within the envelope and other properties of the deep envelope, such as the total stellar mass, rather than by the exact location of the optically thin ``photosphere'' \citep[see \S11.3.1 in][]{2013sse..book.....K}. 
The numerical consequence is that for stars with radiation-diffusion-dominated envelopes, an integration method which obtains the outgoing luminosity (and from it the ``effective'' temperature) need not resolve the photosphere along its steps. 
We will assess whether our code can, as anticipated, produce radiative luminosity comparable to those of a 1D model, even without resolving the photosphere spatially, see in \S\ref{sec_1dflux}.

At a later time, after the merger, the average distance over which the observed radiation forms is substantial. As will be discussed in \S\ref{sec:theplateaustage}, during that epoch, the radiation detected by an observer is typically formed by multiple particles along each ray, creating a widely spread ``effective photosphere'' with an average optical depth between 1 and 10, depending on the specific snapshot.

To summarize, our method does not depend on a hypothetical photosphere being resolved using a small step in optical depth. Instead, it relies on our ability to accurately capture the radiation produced by both optically thin and thick particles.
We will evaluate our method in two scenarios: for optically thick particles representing a stationary radiating mass, and for particles that transition from optically thick to optically thin during dynamical expansion. In the latter, we will assess potential errors in energy output during the transition.

\subsection{Stationary case: a radiative star}
\label{sec_1dflux}

In the binary system that we consider, the donor is convective, and most of the energy transport is provided by convection.
Due to the dominance of the convective energy transport contribution in the total energy transport, neither our radiative energy transport nor radiation flux image processing methods can adequately reproduce the full luminosity of a convective star -- it can only reproduce the part of luminosity carried by the radiation.
Energy losses from the initial convective envelope of a low-mass subgiant during the simulation are insignificant and cannot affect the outcome.
Convective energy transport halts quickly when the ejected material expands, and radiative energy losses become dominant. The only drawback is that using only the radiative flux imaging technique, the obtained imaged luminosity of a binary before the merger is smaller than expected. Hence, in all that follows, we will only show the luminosity curve when the radiative energy exceeds the initial luminosity of the donor. Imaging a convective star can be done separately, using the method described in \citet{2021MNRAS.507..385H}.

Here, we test how our method works when radiative transport dominates. We create a $15\,M_\odot$ zero-age main sequence star in \texttt{MESA} version 10398.
Our 1D model has photospheric radius $R_\text{1D}\simeq4.95\,R_\odot$, luminosity $L\simeq1.98\times10^4\,L_\odot$, and effective temperature $T_\text{eff}=30\,818\,\text{K}$.

We map this 1D model into 3D using \starsmasher, which creates a spherical grouping of particles that are in a close-packed hexagonal configuration.
The goal of this test is to reproduce the surface brightness of a 1D model from a 3D model that has the same thermodynamic properties. To ensure that the 3D model's size and thermodynamic properties are as close as possible to that of the original 1D model, we choose to test our 3D star immediately following the particle mapping without allowing it to relax towards hydrostatic equilibrium. The 3D model has $N\simeq 10^5$ particles and $\mathtt{nnopt}=27$.
We adjust the size of the 3D model such that the particle kernels extend only up to $R_\text{SPH}\equiv \max(r_i + 2h_i) = 4.95\,R_\odot$ and stop the simulation after the first code iteration (after initialization).

Using our image processing method in \S\ref{sec:imaging} we find $L=1.92\times10^4\,L_\odot$
with a spectrum peak temperature of 
$T_\text{sp} = 32\,222\,\text{K}$ (with 95\% of spectral maximum being in the range from 29,000K to  36,250K). 
This is consistent with our method, converging to Equation~(\ref{eq:ldiff_1d}) in the case of a spherically symmetric star. 
It is also consistent with the properties of the radiative layers (including the total luminosity) being independent of the exact boundary (photospheric) values. 
This test demonstrates that unobstructed, optically thick particles create a net radiative flux approximately equal to that of a more exact solution (1D code with photospheric boundary condition).

\subsection{Dynamical case: losses from ejecta}

\begin{figure}
    \centering
    \includegraphics{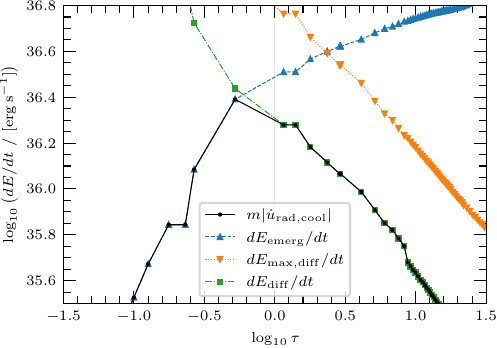}
    \label{fig:fluxtautrack}
    \caption{We show the luminosity of an ejecta particle from simulation \ref{Polytrope} during its transition from the optically thick regime ($\tau \gg 1$) to the optically thin ($\tau \ll 1$). Each data point represents a time iteration in the code at which cooling quantities are updated, which happens in $\sim10\,\text{minute}$ intervals.}
\end{figure}

We consider how a particle radiates while it expands and transitions from an optically thick to an optically thin particle.
We inspect a typical ejecta particle in simulation \ref{Polytrope} (see \S\ref{sec:simulations}) and find that its cooling rate transitions from diffusive $dE_\text{diff}/dt$ in the optically thick regime to emerging $dE_\text{emerg}/dt$ in the optically thin regime rapidly ($\sim10\,\text{minutes}$).
Specifically, a particle would spend $\sim20\,\text{minutes}$ being marginally at a photospheric regime, $\tau\sim1$, and then $10\,\text{minutes}$ later on the next $dE/dt$ update would be optically thin, $\tau\sim0.1$, as shown in Figure~\ref{fig:fluxtautrack}.
The expected maximum cooling rate would be where $dE_\text{emerg}/dt = dE_\text{diff}/dt$. The majority of the observed radiation is expected to originate from particles with their current optical depths ranging between 0.2 and 5.

We can evaluate the error due to the limited time resolution of the cooling updates: the average cooling rate applied during $10\,\text{minutes}$ is $10\,\text{per}\,\text{cent}$ less than it would be at the maximum. 
For the $20\,\text{minute}$ interval when the particle transits from $\tau\sim1$ to $\tau\sim0.1$, the energy it loses by radiation is about $1.1\times 10^{39}\,\text{erg}$.
In comparison, a better time resolution would result in an energy loss of about $1.2\times 10^{39}\,\text{erg}$ during a $20\,\text{minutes}$ transition, which is a fraction of about $0.14$ the total internal energy.

While this specific particle was not caught for the cooling rate calculations at that exact moment, other particles would be.
The closest any of the particles in \ref{Polytrope} get to the maximum cooling rate is $dE_\text{emerg}/dt=1.871\times10^{35}\,\text{erg}\,\text{s}^{-1}$ and $dE_\text{diff}/dt=1.870\times10^{35}\,\text{erg}\,\text{s}^{-1}$.
We estimate that the accuracy of our light curve calculations for a dynamically expanding object due to the limited time updates for cooling rates is within $10\,\text{per\ cent}$ of the fully resolved case.

\subsection{Comparison of methods}\label{sec:medcomparison}
\label{sec:meth_comp}

Here, we review the main assumptions of FLD radiation hydrodynamics as a foundation for comparing our problem-tuned approach with the more versatile method presented in \cite{2007ApJ...667..626K} (hereafter K07).

The FLD method is well-suited for radiative transfer in systems where the diffusion approximation is valid, primarily in optically thick media. However, it has inherent limitations in optically thin regions and in cases where line transport plays a significant role. To address transitions between regimes, K07 employs flux limiters to ensure smooth interpolation between the diffusion and streaming limits. 

The key differences between our FLED approach and the FLD approach of K07 are the choice of hydrodynamic framework and the focus of the simulations. We employ a Lagrangian framework, whereas K07 uses an Eulerian approach. Additionally, our simulations concentrate on slow-ejecta cases ($v/c\ll1$), as opposed to scenarios like supernovae. The main update we do is only for the radiation energy density, and this update is derived from the formal radiative transfer equation that is treated as time-independent. This simplification is justified by the significantly slower evolution of matter in the system compared to the radiation energy density. Consequently, we compute the radiation energy density change in an optically thin regime using a quasi-stationary solution at each time step.

\begin{enumerate}

\item {\bf Local Thermodynamic Equilibrium (LTE)}: In both our approach and K07, the radiation field is assumed to behave as a blackbody, simplifying the spectral treatment. This implies that both methods assume LTE, which is valid in optically thick regions but introduces limitations in optically thin regions where non-thermal effects may dominate.

\item {\bf Momentum Equation}: K07 incorporates the feedback of radiation on the momentum equation. This feedback is important in the dynamic diffusion regime, where terms proportional to $v/c$ contribute significantly. However, their momentum equation is inherently non-conserving, as FLD does not explicitly track momentum transfer between the gas and radiation fields. In our case, due to the small velocities of the ejecta, we chose not to include radiative feedback. Post-run evaluations show that radiative acceleration is small compared to local gravitational acceleration in our simulations. However, future extensions of our method could incorporate radiative acceleration for completeness.

\item \textbf{Opacities:}
K07 introduces four types of opacities:
\begin{itemize}
    \item Planck opacity ($\kappa_P$): Used to characterize the material's ability to emit radiation in the form of blackbody radiation.
    \item Rosseland opacity ($\kappa_R$): Used to compute the mean opacity for radiative diffusion in optically thick regions.
    \item Energy opacity ($\kappa_E$): Used to determine the energy absorption from the radiation field.
    \item Flux opacity ($\kappa_F$): Used to calculate the rate of radiative emission of the fluid.
\end{itemize}

In K07, these opacities are defined as linear absorption coefficients, while in our approach, they are expressed as mass absorption coefficients. The two are related by $\kappa^{\text{K07}} = \rho \kappa^{\text{HI}}$, where $\rho$ is the density of the medium.

K07 assumes $\kappa_P = \kappa_E$ under the assumption of blackbody radiation. To optimize the accuracy for optically thick flows, they set $\kappa_F = \kappa_R$. In our simulations, the values of Planck and Rosseland opacities are nearly identical in the optically thick regime. 

Currently, we use the same opacity for both radiation absorption and emission. However, it is possible to refine this approach by using different opacities for these processes, depending on the local optical depth. This flexibility allows us to better account for radiative processes in both optically thick and thin regimes.
For optically thin regions, switching to Planck opacities in the future can improve the accuracy of radiative emission calculations.

\item \textbf{Optically Thick Regime:}
In both approaches, the radiative flux in optically thick regions is modeled using Fick's law under the diffusion approximation. For $\tau \gg 1$, K07 describes the flux as:

\begin{equation}
F \to -\frac{c}{3 \kappa_R^{\text{K07}}} \nabla U_\text{rad},
\end{equation}

which is equivalent to our formulation for diffusive flux in an optically thick regime:

\begin{equation}
F_{\rm diff} = -\frac{c}{3 \kappa_R^{\text{HI}}\rho } \nabla U_\text{rad}.
\end{equation}.

\item \textbf{Empty Space Regime:}  
When there is no matter, the radiative flux is not altered along its path. This is known as the free-streaming regime.  
In K07, the free-streaming regime is assumed for $\tau \ll 1$, where photons move freely at the speed of light ($c$) without significant interactions with the medium. In this regime, the radiative flux is approximated as:  
\begin{equation}
\mathbf{F} \to c U_\text{rad} \hat{n},
\end{equation} 
where $U_\text{rad}$ is the radiation energy density, and $\hat{n}$ is the direction of the flux.

In the absence of interactions with matter (e.g., outside SPH particles), our Lagrangian framework naturally ensures that the flux is conserved. However, we cannot fully recover the same free-streaming behavior as in K07 in the following meaning: we do not evolve radiation (not we recover flux value) between matter during the simulations. Radiation leaving the domain with matter into empty space and passing through regions with no opacity at all does so immediately. The consequence of this limitation arises not during simulations but during light-curve image processing. Specifically, radiation emitted from particles located at $10 R_\odot$ and $300 R_\odot$ will be registered by our imaginary observer immediately, leading to a slightly shorter duration of the event's expansion phase. The maximum error in the light curve in our case is estimated to be about 10 minutes. Accounting for the time it takes light to travel to the same location can be incorporated into the image processing in the future.

Please note that the actual escaping flux during the simulation in both methods depends on the treatment of the intermediate regime, as discussed below in the context of the method comparison. Nonetheless, our image processing accurately recovers the correct flux.

\item \textbf{Transition between Optically Thin and Thick Regimes:}
For $\tau \ll 1$, the K07 approach naturally transitions to the free-streaming regime. K07 employs a flux limiter, $\lambda$ \citep{1981ApJ...248..321L}, which provides a smooth numerical transition between the diffusion limit and the free-streaming limit. This technique ensures that the approach recovers the optically thick regime in one limit and the free-streaming regime in the other.

In contrast, our approach uses a $\tau$-dependent solution to the radiative energy transport equation, specifically tailored for the regime of small but finite optical depth.
Our setup ensures that radiation emitted from optically thin particles is additional to the radiation passing through the medium. For small $\tau$, this approach naturally sums the radiation contributions from every particle with small optical depth along the ray path. 

On the other hand, in our implementation, radiation produced in the diffusion regime can be absorbed only by the first, usually most optically thick, particle along the ray during the dynamical simulation. Subsequent particles along the ray contribute their emission to the flux but do not absorb energy from the main flux during the simulation. 

\end{enumerate}

An analytical comparison of the methods is not feasible, so we perform a numerical experiment for a spherically symmetric case, using a 1D grid. This experiment uses the same assumptions and limitations applied in both methods, comparing their results to the exact numerical solution.

\begin{figure}
    \centering
    \includegraphics[width=\columnwidth]{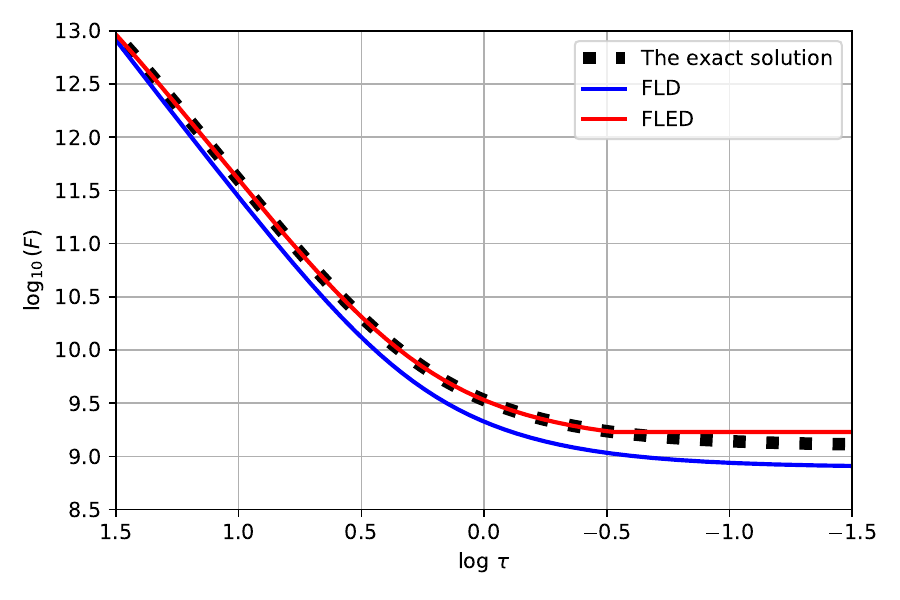}
    \label{fig:flux_comparison}
    \caption{The comparison of radiative fluxes obtained using the exact numerical solution of the radiative transfer equation near $\tau = 1$, the FLD method with a $\lambda$-limiter, and the FLED method described in this paper. The radiative fluxes are computed for the following simplified 'adiabatic' profile: $\rho = \rho_0 (r/r_0)^{-3}$, $T(r) = T_0 (\rho / \rho_0)^{\gamma - 1}$, with $\rho_0 = 10^{-8}~\mathrm{g/cm^3}$, $T_0 = 10^6~\mathrm{K}$, $r_0 = 2R_\odot$, $r_{\rm final} = 100 r_0$, $\kappa = 1~\mathrm{cm^2/g}$, and $\gamma = 5/3$.
}
\end{figure}

For any given temperature, density, and opacity profile, the exact solution to the time-independent (static) radiative transfer equation can be computed for both inward and outward intensities. These intensities allow the calculation of the radiative flux, as illustrated in Figure~\ref{fig:flux_comparison}. This is referred to as the exact solution. Additionally, the intensities in the exact solution establish a static radiative energy distribution. It is important to note that this distribution differs significantly from the radiative energy distribution arising solely from matter in optically thin regions, where the passing radiative flux plays a dominant role.

The FLD method can be applied to static diffusion cases, where the static FLD flux is determined using the static radiation energy distribution. As shown in Figure~\ref{fig:flux_comparison}, the free-streaming radiative flux in the FLD method is, as expected, smaller than that predicted by the exact solution. Numerical experiments indicate that this escaping flux in the FLD method is approximately equivalent to the flux obtained through direct integration of the radiative transfer equation when no emission occurs in regions with $\tau \leq 3$.

Empirically, our simulations reveal that the last particle to absorb outgoing radiation is typically located at an optical depth of $\tau = 0.3$. We computed the FLED flux on the same 1D mesh grid as described above -- distinct from the SPH realization where we do not trace the flux in any case -- but using the same criteria outlined in \S~\ref{sec:radiativecooling}. This involved comparing local optically thin emission with diffusion-driven emission, selecting the appropriate regime for each point, and assuming no absorption (no passing flux attenuation) occurs in the region $\tau \leq 0.3$. The resulting FLED flux is depicted in Figure~\ref{fig:flux_comparison}. In this scenario, less radiative energy is absorbed by optically thin particles during the simulation. However, the flux derived from image processing incorporates all contributions.

Qualitatively, the standard FLD method retains more radiation in the optically intermediate layers and allows less radiation to escape. In contrast, our FLED method permits more flux to escape, retaining less energy in the optically thin particles. For the example shown, the escaping FLD flux is approximately $62\%$ of the flux in the exact solution, while the escaping flux in our method is $133\%$ of the exact solution. The difference between the FLD method and the exact solution is not constant: when an opacity drop is introduced, linearly transitioning from $1~\mathrm{cm^2/g}$ to $0.01~\mathrm{cm^2/g}$ in the region from $6000~\mathrm{K}$ to $4000~\mathrm{K}$, the FLD flux decreases further to $48\%$ of the exact flux at infinity.

In our method, the amount of escaping flux remains consistent, but can be further refined in the future by enabling more particles along the ray to absorb radiation. This improvement would increase physical accuracy at the cost of higher computational complexity.

\begin{table}
    \centering%
    \setlength{\tabcolsep}{0.4\tabcolsep}%
    \caption{Our relaxed 3D \starsmasher stellar models.  They are denoted as ``Donor'' or ``Companion'' depending on its use in our dynamical simulations shown in Table~\ref{tab:simulations}. Mass $M$ is in $M_\odot$. For stars built from an initial 1D \texttt{MESA} model, we show the \texttt{MESA} photospheric radius $R_\text{1D}$ in $R_\odot$. The SPH radius $R_\text{SPH}=\max(r_i + 2h_i)$ and the maximum particle radial position $R_\text{SPH}^*\equiv \max(r_i)$ are shown in $R_\odot$. For point mass companions we show $R_\text{SPH}$ as the gravity softening radius of the point mass. Each simulation consists of $N$ particles.}%
    \vspace{-0.1in}
    \label{tab:stars}%
    \begin{tabular}{lllrrrrr}
        \tabletophlineone
        Model & Star & Type & $M$ & $R_\mathrm{1D}$ & $R_\mathrm{SPH}$ & $R_\mathrm{SPH}^*$ & $N$ \\[0.04in]
        \tabletophlinetwo
         $\text{M15R37}\simulationlabel{M15R37}{relax:M15R37}$ & Donor & \texttt{MESA} & 1.52 & 3.715 & 4.062 & 3.565 & 99955\\
        $\text{M15R37G}\simulationlabel{M15R37G}{relax:M15R37G}$ & Donor & \texttt{MESA} & 1.52 & 3.715 & 3.986 & 3.574 & 299929\\
        $\text{M15R40}\simulationlabel{M15R40}{relax:M15R40}$ & Donor & \texttt{MESA} & 1.52 & 3.990 & 4.299 & 3.904 & 99955\\
        \ \\
        $\text{M02R02poly}\simulationlabel{M02R02poly}{relax:M02R02poly}$ & Companion & \texttt{SPH} & 0.16 & -- & 0.209 & 0.141 & 1038\\
        $\text{M01R01}\simulationlabel{M01R01}{relax:M01R01}$ & Companion & \texttt{MESA} & 0.10 & 0.122 & 0.149 & 0.101 & 1038\\
        $\text{M02R02}\simulationlabel{M02R02}{relax:M02R02}$ & Companion & \texttt{MESA} & 0.16 & 0.182 & 0.222 & 0.150 & 1038\\
        $\text{M02R02pm}\simulationlabel{M02R02pm}{relax:M02R02pm}$ & Companion & \texttt{SPH} & 0.16 & -- & 0.182 & -- & 1\\
        \hline
    \end{tabular}
\end{table}

\begin{table*}
    \centering
    \setlength{\tabcolsep}{0.5\tabcolsep}
    \caption{We show our \starsmasher dynamical simulations with their corresponding original donor and companion stars from Table~\ref{tab:stars}. 
    The initial orbital period $P_0$ is shown in days, the donor radius $R_{\text{1D},d}$ and companion radius $R_{\text{1D},c}$ are in $R_\odot$ and are taken from the initial 1D \texttt{MESA} models (the same as $R$ in Table~\ref{tab:stars}), and the SPH zero-density radii of our companion stars $R_{\text{SPH},c} = \max(r_i+2h_i)$ is in $R_\odot$, where in the case of a point mass $R_{\text{SPH},c}$ is the gravity softening radius. We calculate the fraction of the Roche lobes that are filled at the start of each simulation for the donor $f_{\text{RLOF},d}$ and companion $f_{\text{RLOF},c}$. The opacities $\kappa$ correspond to our three opacity lookup tables defined in \S\ref{sec:opacities}. The final simulation times $t_f$ and plunge times $t_\text{plunge}$ are given in days. The ejected mass $M_\text{ej}$ is given in $M_\odot$ and corresponds with the value measured at $t-t_\text{plunge}=10\,\text{days}$.}
    \vspace{-0.1in}
    \label{tab:simulations}
    \begin{tabular}{lrrrrrrrrrrrrrr}
        \tabletophlineone
        Model & Donor & Companion & $P_0$ & Cooling & Heating? & $R_{\text{1D},d}$ & $R_{\text{1D},c}$ & $R_{\text{SPH},c}$ & $f_{\text{RLOF},d}$ & $f_{\text{RLOF},c}$ & $\kappa$ & $t_f$ & $t_\text{plunge}$ & $M_\text{ej}$ \\[0.04in]
        \tabletophlinetwo
        $\text{LargeD}\simulationlabel{LargeD}{d086c}\simulationlabel{LargeD}{LargeD}$ & \ref{relax:M15R40} & \ref{relax:M02R02} & 1.66 & Dense & Yes & 3.990 & 0.182 & 0.222 & 0.977 & 0.101 & \ref{fullopacities} & 54.34 & 20.81 & 0.067\\
        $\text{LargeF}\simulationlabel{LargeF}{d085c}\simulationlabel{LargeF}{LargeF}$ & \ref{relax:M15R40} & \ref{relax:M02R02} & 1.66 & Fluffy & Yes & 3.990 & 0.182 & 0.222 & 0.977 & 0.101 & \ref{fullopacities} & 70.48 & 27.21 & 0.070\\
        $\text{MainDtiny}\simulationlabel{MainDtiny}{d080c}\simulationlabel{MainDtiny}{MainDtiny}$ & \ref{relax:M15R37} & \ref{relax:M01R01} & 1.47 & Dense & Yes & 3.715 & 0.122 & 0.149 & 0.974 & 0.088 & \ref{fullopacities} & 48.40 & 12.95 & 0.042\\
        $\text{MainD}\simulationlabel{MainD}{d082c}\simulationlabel{MainD}{MainD}$ & \ref{relax:M15R37} & \ref{relax:M02R02} & 1.55 & Dense & Yes & 3.715 & 0.182 & 0.222 & 0.980 & 0.108 & \ref{fullopacities} & 73.04 & 13.70 & 0.061\\
        $\text{MainFP}\simulationlabel{MainFP}{d090c}\simulationlabel{MainFP}{MainFP}$ & \ref{relax:M15R37} & \ref{relax:M02R02} & 1.55 & Fluffy & Yes & 3.715 & 0.182 & 0.222 & 0.980 & 0.108 & \ref{planckonlyopacities} & 399.87 & 13.67 & 0.081\\
        $\text{MainFR}\simulationlabel{MainFR}{d113c}\simulationlabel{MainFR}{MainFR}$ & \ref{relax:M15R37} & \ref{relax:M02R02} & 1.55 & Fluffy & Yes & 3.715 & 0.182 & 0.222 & 0.980 & 0.108 & \ref{rosselandonlyopacities} & 399.87 & 13.74 & 0.071\\
        $\text{MainF\ Grand}\simulationlabel{MainF\ Grand}{d106c}\simulationlabel{MainF\ Grand}{MainFGrand}$ & \ref{relax:M15R37G} & \ref{relax:M02R02} & 1.57 & Fluffy & Yes & 3.715 & 0.182 & 0.222 & 0.978 & 0.106 & \ref{fullopacities} & 44.80 & 19.33 & 0.058\\
        $\text{MainFtiny}\simulationlabel{MainFtiny}{d079c}\simulationlabel{MainFtiny}{MainFtiny}$ & \ref{relax:M15R37} & \ref{relax:M01R01} & 1.47 & Fluffy & Yes & 3.715 & 0.122 & 0.149 & 0.974 & 0.088 & \ref{fullopacities} & 74.53 & 12.98 & 0.057\\
        $\text{MainF}\simulationlabel{MainF}{d081c}\simulationlabel{MainF}{MainF}$ & \ref{relax:M15R37} & \ref{relax:M02R02} & 1.55 & Fluffy & Yes & 3.715 & 0.182 & 0.222 & 0.980 & 0.108 & \ref{fullopacities} & 109.01 & 13.74 & 0.062\\
        $\text{PMD\ No\ Heat}\simulationlabel{PMD\ No\ Heat}{d094c}\simulationlabel{PMD\ No\ Heat}{PMDNoHeat}$ & \ref{relax:M15R37} & \ref{relax:M02R02pm} & 1.55 & Dense & No & 3.715 & -- & -- & 0.979 & -- & \ref{fullopacities} & 25.45 & 14.00 & 0.079\\
        $\text{PMD}\simulationlabel{PMD}{d095c}\simulationlabel{PMD}{PMD}$ & \ref{relax:M15R37} & \ref{relax:M02R02pm} & 1.55 & Dense & Yes & 3.715 & -- & -- & 0.979 & -- & \ref{fullopacities} & 34.80 & 20.97 & 0.081\\
        $\text{PM\ No\ Heat}\simulationlabel{PM\ No\ Heat}{d093c}\simulationlabel{PM\ No\ Heat}{PMNoHeat}$ & \ref{relax:M15R37} & \ref{relax:M02R02pm} & 1.55 & Fluffy & No & 3.715 & -- & -- & 0.979 & -- & \ref{fullopacities} & 65.40 & 17.02 & 0.089\\
        $\text{PM}\simulationlabel{PM}{d089c}\simulationlabel{PM}{PM}$ & \ref{relax:M15R37} & \ref{relax:M02R02pm} & 1.55 & Fluffy & Yes & 3.715 & -- & -- & 0.979 & -- & \ref{fullopacities} & 184.42 & 18.61 & 0.079\\
        $\text{Polytrope}\simulationlabel{Polytrope}{d096c}\simulationlabel{Polytrope}{Polytrope}$ & \ref{relax:M15R37} & \ref{relax:M02R02poly} & 1.55 & Fluffy & Yes & 3.715 & -- & 0.209 & 0.979 & 0.102 & \ref{fullopacities} & 71.80 & 13.65 & 0.065\\
        \hline
    \end{tabular}
\end{table*}

\section{Simulations}\label{sec:simulations}
Our base model binary system is a $1.52\,M_\odot$ donor star with a $0.16\,M_\odot$ companion.
The pairing of $1.52\,M_\odot$ and $0.16\,M_\odot$ stars at the orbital period of $P\approx 1.4$ days is one of the preferred suggested V1309~Sco progenitor configurations \citep{2011A&A...531A..18S}.
This period suggests the donor star has a radius of about $3.7R_\odot$.

We do not vary the mass of the donor star, but we do explore the effect of the donor's radius.
Specifically, we explore the donors with $3.7$ and $4\,R_\odot$.
The exact parameters of the initial 1D {\tt MESA} stellar models from which the donor and companion stars are created are provided in Table~\ref{tab:stars}.

For each donor star that originates from a \texttt{MESA} profile, we find the 3D \starsmasher relaxation which best matches the total energy profile of the \texttt{MESA} model in the top $0.1\,M_\odot$ by varying the \texttt{nnopt} input parameter in \starsmasher \nopagebreak[4]\citep{2021MNRAS.507..385H}.
We find a match of $\gtrsim 99.9\,\text{per\,cent}$ with $\mathtt{nnopt}=16$ in both the $3.401\,R_\odot$ and $3.990\,R_\odot$ donor, and our $3.715\,R_\odot$ donor is the same as in \citet{2021MNRAS.507..385H}.

To explore the effect of the companion mass we use $0.1\,M_\odot$ and $0.16\,M_\odot$ companions.
Our companion stars are either a relaxed 3D model built from a 1D \texttt{MESA} star, a relaxed 3D polytrope with polytropic index $n=3/2$, or a point mass represented by a single SPH particle.
In the case of a point mass, the softening length is $0.182\,R_\odot$.

We continued the simulations until at least the moment of dust formation (see more details in \S\ref{sec:results_stages}), with several evolving longer.
We provide the parameters for our \starsmasher dynamical simulations in Table~\ref{tab:simulations}.

\begin{table}
    \begin{center}
    \setlength{\tabcolsep}{7pt}
    \caption{We show the results of our simulations in Table~\ref{tab:simulations} where the plunge-in time $t_\text{plunge}$ is in days. We show the ejected mass $M_\text{ej}$ measured at time $t-t_\text{plunge}=10\,\text{days}$. For select quantities we calculate the absolute relative fractional difference to a baseline simulation (denoted by \text{\textdagger}) as defined in Equation~(\ref{eq:comparisondelta}).}
    \label{tab:comparisons} 
    \vspace{-0.1in}
    \begin{tabular}{lrrrr}
        \tabletophlineone
        Model & $t_\text{plunge}$ & $\Delta t_\text{plunge}$ & $M_\text{ej}$ & $\Delta M_\text{ej}$ \\[0.04in]
        \tabletophlinetwo
        \textbf{Donor size} \\
        \ref{MainF}\textsuperscript{\textdagger} & 13.74 &  & 0.062 & \\
        \ref{MainFGrand} & 19.33 & $ 40.7\%$ & 0.058 & $ -7.1\%$\\
        \ref{MainD} & 13.70 & $ -0.3\%$ & 0.061 & $ -2.2\%$\\
        \ref{LargeF} & 27.21 & $ 98.0\%$ & 0.070 & $ 11.5\%$\\
        \ref{LargeD} & 20.81 & $ 51.4\%$ & 0.067 & $  7.7\%$\\
        \\
        \textbf{Companion type} \\
        \ref{MainF}\textsuperscript{\textdagger} & 13.74 &  & 0.062 & \\
        \ref{Polytrope} & 13.65 & $ -0.7\%$ & 0.065 & $  4.5\%$\\
        \ref{PM} & 18.61 & $ 35.4\%$ & 0.079 & $ 27.3\%$\\
        \\
        \textbf{Companion mass} \\
        \ref{MainF}\textsuperscript{\textdagger} & 13.74 &  & 0.062 & \\
        \ref{MainD} & 13.70 & $ -0.3\%$ & 0.061 & $ -2.2\%$\\
        \ref{MainFtiny} & 12.99 & $ -5.5\%$ & 0.057 & $ -9.2\%$\\
        \ref{MainDtiny} & 12.95 & $ -5.8\%$ & 0.042 & $-32.6\%$\\
        \\
        \textbf{Heating} \\
        \ref{PM}\textsuperscript{\textdagger} & 18.61 &  & 0.079 & \\
        \ref{PMD} & 20.97 & $ 12.7\%$ & 0.081 & $  1.9\%$\\
        \ref{PMNoHeat} & 17.02 & $ -8.5\%$ & 0.089 & $ 12.0\%$\\
        \ref{PMDNoHeat} & 14.00 & $-24.8\%$ & 0.079 & $  0.0\%$\\
        \\
        \textbf{Opacities} \\
        \ref{MainF}\textsuperscript{\textdagger} & 13.74 &  & 0.062 & \\
        \ref{MainFP} & 13.67 & $ -0.5\%$ & 0.081 & $ 29.9\%$\\
        \ref{MainFR} & 13.76 & $  0.1\%$ & 0.071 & $ 13.9\%$\\
        \hline
    \end{tabular}
    \end{center}
    \vspace{-0.08in}
    {\textsuperscript{\textdagger}}The baseline simulation used for absolute relative fractional differences $\Delta$ as defined in Equation~(\ref{eq:comparisondelta}).
\end{table}

\subsection{Comparison groups}

We make the following comparisons, as shown in Table~\ref{tab:comparisons}: \label{enum:comparisons}

\vspace{\itemsep}

\noindent\textit{\textbf{Cooling type.}} To understand how our fluffy and dense cooling regimes differ across a wide parameter space, we include tests in each of the donor size, companion mass, and heating comparisons.

\vspace{\itemsep}

\noindent\textit{\textbf{Donor size.}} We create donors of two different sizes: \ref{relax:M15R37}, and \ref{relax:M15R40}, as shown in Table~\ref{tab:stars}.
Each donor is placed in a binary with the same companion, \ref{relax:M02R02}.
We create one set of simulations with fluffy cooling and heating turned on and a second, identical set but with dense cooling instead of fluffy.

\vspace{\itemsep}

\noindent\textit{\textbf{Companion type.}} The accretor is one of \ref{relax:M02R02pm}, \ref{relax:M02R02}, or \ref{relax:M02R02poly}, each with of mass $0.16\,M_\odot$ and in the fluffy cooling regime with heating turned on.

\vspace{\itemsep}

\noindent\textit{\textbf{Companion mass.}} 
The companion is either \ref{relax:M01R01} ($0.1\,M_\odot$) or \ref{relax:M02R02} ($0.16\,M_\odot$) and both fluffy and dense cooling regimes are tested with heating turned on.

\vspace{\itemsep}

\noindent\textit{\textbf{Heating.}} We investigate the effect of the radiative heating using simulations with a $0.16\,M_\odot$ point mass companion, in both fluffy and dense cooling regimes.

\vspace{\itemsep}

\noindent\textit{\textbf{Opacities.}} We check the effect of opacities by using three different opacity tables, as shown in Figure~\ref{fig:opacities}: \ref{MainFP}, with Rosseland opacities excluded (\ref{planckonlyopacities}); \ref{MainFR}, with Planck opacities excluded (\ref{rosselandonlyopacities}); and \ref{MainF}, with Planck opacities excluded (\ref{fullopacities}), as the opacities for $T<3\,000\,\text{K}$ (which includes the Planck opacities temperature regime) are manually set to $10^{-4}\,\text{cm}^2\,\text{g}^{-1}$.
Only fluffy cooling is used with heating enabled.

\subsection{Key quantities}\label{sec:keyquantities}

When we describe a simulation, we use the following quantities and their definitions: \label{enum:quantity}

\vspace{\itemsep}

\noindent\textit{\textbf{Plunge-in time.}} The time of the plunge-in event $t_\text{plunge}$ is the first time step at which more than $0.05\,M_\odot$ of the companion's mass is within a distance of $R_\text{SPH}$ from the donor's center of mass.

\vspace{\itemsep}
    
\noindent\textit{\textbf{Ejected mass.}} A particle is identified as a member of the ejecta when its total energy $E_\text{tot} \geq 0$.
We measure the total mass of the ejected particles $M_\text{ej}$ at $t-t_\text{plunge}=10\,\text{days}$.

\vspace{\itemsep}

\noindent\textit{\textbf{Expansion velocity}} The flux-weighted average of the expansion velocity $\langle v_\text{exp}\rangle$ from Equation~(\ref{eq:averagevexp}) represents a synthetic signal which can be compared to the observed real expansion velocity.

\vspace{\itemsep}
    
\noindent\textit{\textbf{Cooling luminosity.}} We define the cooling luminosity as
\begin{align}\label{eq:Lcool}
    L_\text{cool} \equiv \sum_i^N |\dot{u}_{\text{rad,cool},i}|m_i - \sum_i^N |\dot{u}_{\text{rad,heat},i}|m_i
\end{align}
where simulations without heating have $\dot{u}_{\text{rad,heat},i}=0$.

\vspace{\itemsep}
    
\noindent\textit{\textbf{Observed luminosity.}} The method with which the ``observed'' luminosity $L_{\rm obs}$ is obtained is described in \S\ref{sec:imaging}.
The assumptions about dust opacity present an additional uncertainty in the generation of the observed light curve.
The overall duration of the event is very rapid, several dozens of days.
It is not clear yet if dust would be able to form.
By default we assume that dust cannot be formed during the simulations and apply opacity table \ref{fullopacities}.
Only simulations \ref{MainFP} and \ref{MainFR} have full dust-containing opacity tables, \ref{planckonlyopacities} and \ref{rosselandonlyopacities} respectively.
To understand the effect that dust has on the observed light curve, we also consider during the image processing of all other simulations that particles with $100\,\text{K} < T < 1\,000\,\text{K}$ have opacities no smaller than $\kappa_\text{dust}=1\,\text{cm}^2\,\text{g}^{-1}$.

\vspace{\itemsep}

To compare a single quantity between simulations we define a ``baseline'' simulation for each of our comparisons and obtain the absolute relative difference of any quantity $A$ to the baseline value $A_\text{base}$:
\begin{align}
    \Delta A\equiv \frac{A - A_\text{base}}{A_\text{base}}. \label{eq:comparisondelta}
\end{align}

\begin{figure*}
    \centering
    \includegraphics{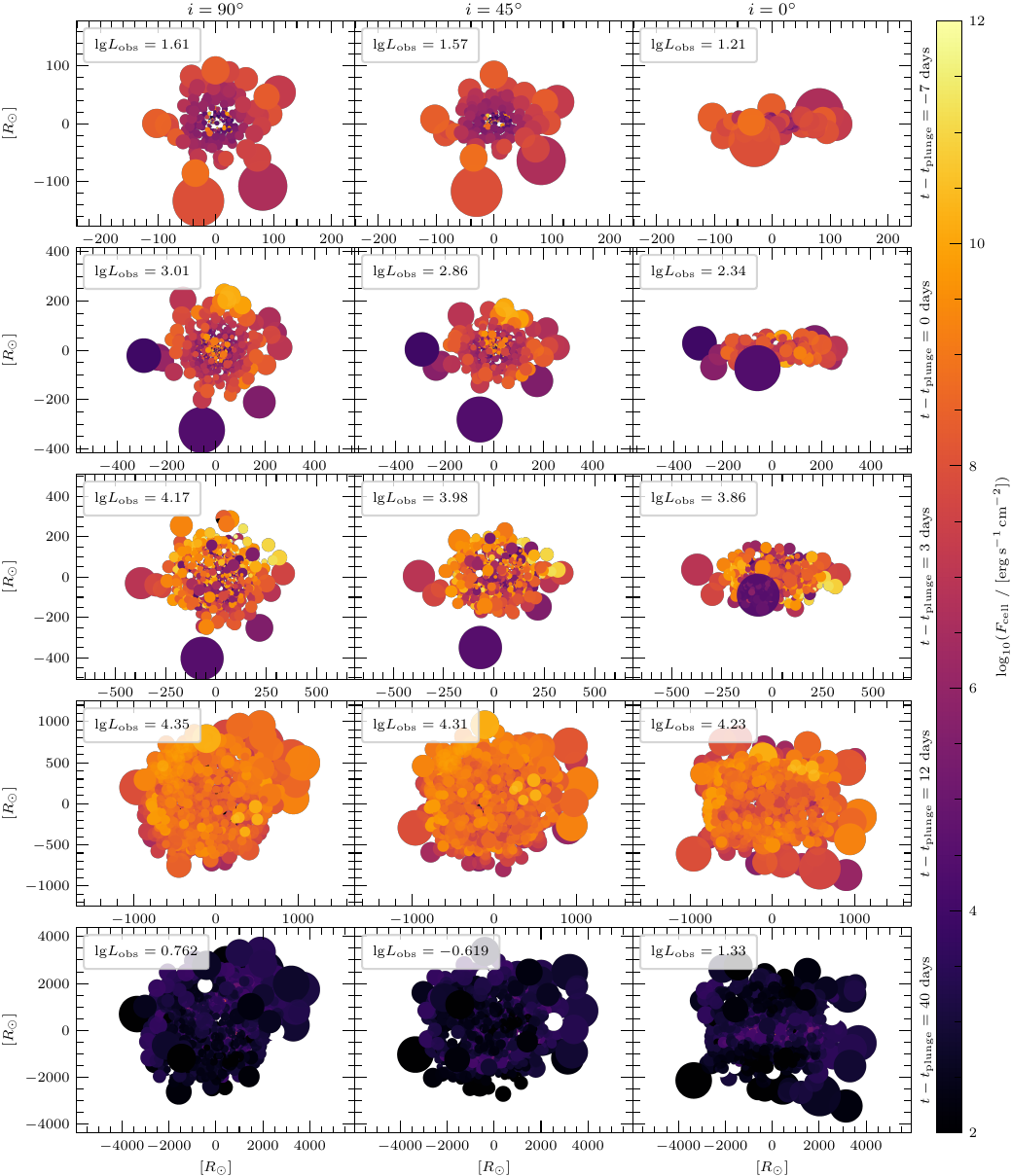}
    \label{fig:Fcellimagesangles}
    \caption{We show for \ref{MainF} the observed flux $F_\text{cell}$ from our imaging method in \S\ref{sec:imaging} at various times $t$ after the plunge-in at $t_\text{plunge}$ from an inclination angle $i=90^\circ$ (left; ``top down'' view on the orbital plane), $i=45^\circ$ (middle), and $i=0^\circ$ (right; ``edge on'' view). Each image in this figure shows the full simulation domain, has a resolution of $2288\times1688$, the quoted values of observed luminosity $\text{lg}L_\text{obs}$ are the  values obtained using our adaptive resolution, as in Figure~\ref{fig:main_light_curve}. Each panel spans the $xy$ ``screen space'' plane in $R_\odot$ (see \S\ref{sec:imaging}). From the top to bottom, the panel rows show samples from the evolutionary stages described in \S\ref{sec:results_stages}: pre-plunge, plunge, hot peak, plateau, and dimming stages. Here we use constant dust opacities $\kappa_\text{dust}=1\,\text{cm}^2\,\text{g}^{-1}$ for particles with $100<T_i<1\,000\,\text{K}$ and only particles with $\tau_i>10^{-5}$ are shown.}
\end{figure*}

\begin{figure}
    \centering
    \includegraphics{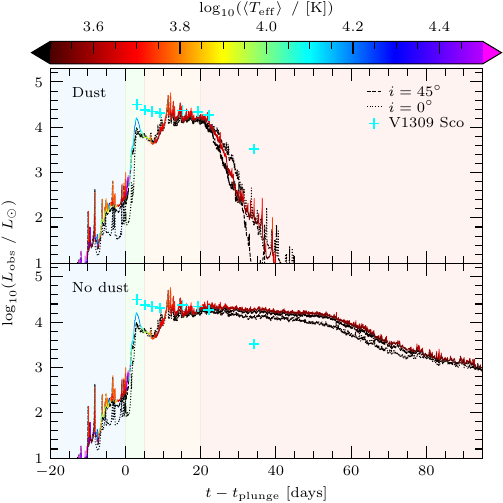}
    \label{fig:main_light_curve}
    \caption{The luminosity $L_\text{obs}$ as measured by an infinitely far away observer for simulation \ref{MainF} using the method described in \S\ref{sec:imaging}. In the top panel, dust opacities $\kappa_\text{dust} = 1\,\text{cm}^2\,\text{g}^{-1}$ are used for particles with $100 < T_i < 1\,000\,\text{K}$. Our standard $\kappa_\text{dust} = 10^{-4}\,\text{cm}^2\,\text{g}^{-1}$ dust opacities are used in the bottom panel for particles with $100 < T_i < 1\,000\,\text{K}$. The $i=90^\circ$ (``top-down'' view on the orbital plane) light curves are coloured using the spectrum peak temperatures. The $i=45^\circ$ (dotted black lines) and  $i=0^\circ$ (dashed black lines; ``edge on'' view) are not colored using the spectrum peak temperatures. We show the V1309~Sco (cyan crosses) luminosities $L$ from table~1 of \citet{2011A&A...528A.114T}, shifted in time so that the maximum value of $L$ in that table corresponds with the maximum $L_\text{obs}$ during the hot peak in \ref{MainF}. The colors of the V1309~Sco data are only aesthetic and do not represent $T_\text{eff}$. The \ref{MainF} evolutionary stages are shown as light blue for pre-plunge, light green for hot peak, light orange for plateau, and light red for dimming.}
\end{figure}

\begin{figure*}
    \centering
    \includegraphics[width=3.3in]{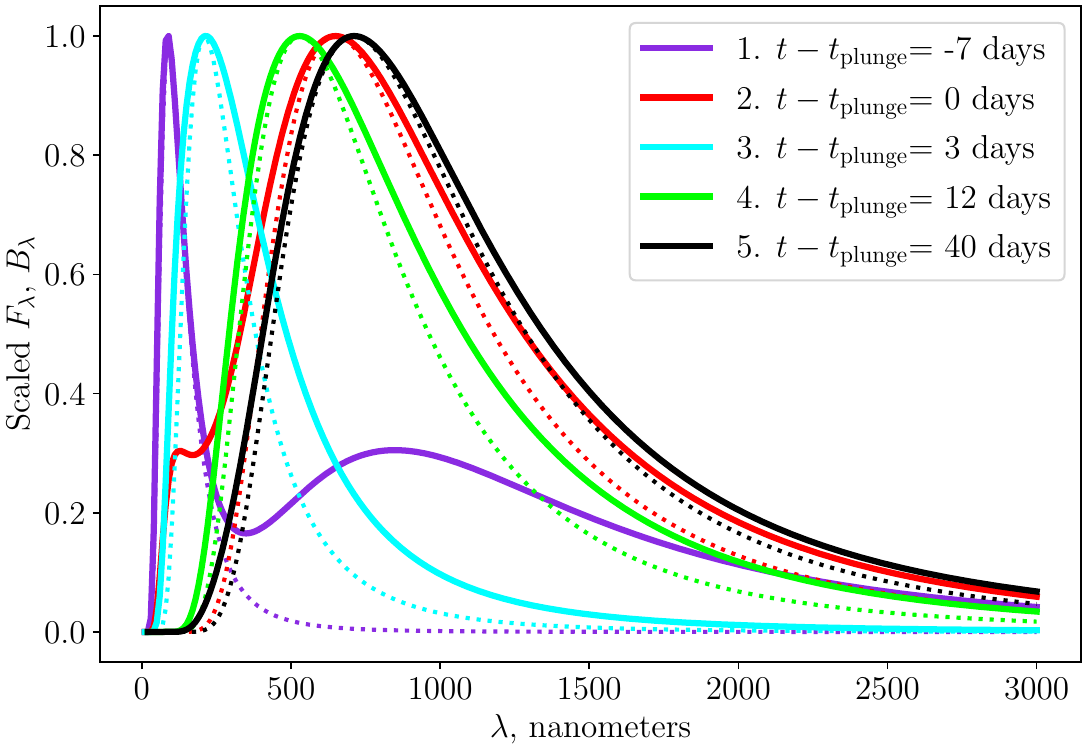}
    \includegraphics[width=3.3in]{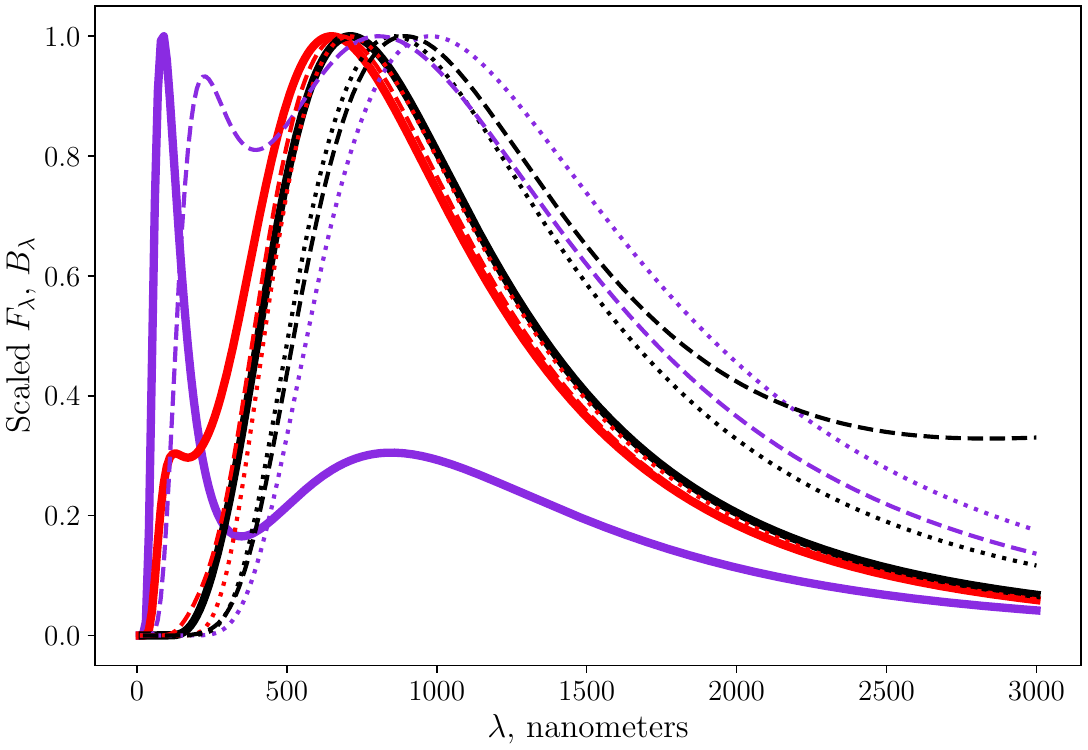}
    \label{fig:spectrums}
    \caption{We show the spectral distribution in observed flux from our imaging method in \S\ref{sec:imaging} at various times $t-t_\text{plunge}$ in simulation \ref{MainF} from an inclination angle $i=90^\circ$ compared to the blackbody (left panel) and for three inclination angles (right panel). In the left panel:
    pre-plunge (violet), plunge (blue), hot peak (green), plateau (red), and dimming (black). The solid line shows the integrated spectral distribution, and the dotted lines show the blackbody distribution. In the right panel, we only select the stages where angular dependence is observed: pre-plunge (violet), plunge (blue), and dimming (black). There, solid lines are for cases of  $i=90^\circ$ (``top down'' view on the orbital plane), dashed lines are for cases of $45^\circ$, and dotted lines show cases of $0^\circ$ (``edge on'' view). 
    The spectrum peak temperatures for the left panel are: 1) $32\,200\,\text{K}\pm_{3\,000\,\text{K}}^{4\,200\,\text{K}}$ 
    (with the secondary peak at $3\,410\,\text{K}\pm_{600\,\text{K}}^{670\,\text{K}}$), 
    2) $4\,460\,\text{K}\pm_{700\,\text{K}}^{810\,\text{K}}$, 
    3) $13\,810\,\text{K}\pm_{2\,650\,\text{K}}^{3\,250\,\text{K}}$, 4) $5\,470\,\text{K}\pm_{870\,\text{K}}^{970\,\text{K}}$, and 5) $4\,080\,\text{K}\pm_{590\,\text{K}}^{670\,\text{K}}$. Each spectral distribution is scaled to its maximum value. A corresponding blackbody spectrum is shown in the left panel peaks at the identified peak in the spectral distribution (dotted lines of the same color). The spectral distributions are obtained assuming that there are dust opacities and ensuring spatial resolution convergence of the object.}
\end{figure*}

\begin{figure}
    \centering
    \includegraphics{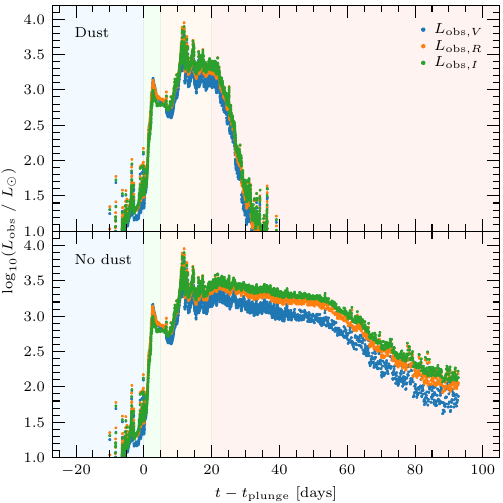}
    \label{fig:MainF_Lv_Lr_Li}
    \caption{We show observed luminosities $L_{\text{obs}, V}$ (blue), $L_{\text{obs},R}$ (orange), and $L_{\text{obs},I}$ (green) in the $V$-band, $R$-band, and $I$-band respectively for simulation \ref{MainF}, as described in \S\ref{sec:imaging} (Equation~\ref{eq:Lobs_spectra}). In the top panel we use dust opacities $\kappa_\text{dust}=1\,\text{cm}^2\,\text{g}^{-1}$ and in the bottom panel we do not use dust opacities. The \ref{MainF} evolutionary stages are shown as light blue for pre-plunge, light green for hot peak, light orange for plateau, and light red for dimming.}
\end{figure}

\section{Results: the main stages}\label{sec:results_stages}

\label{sec6_main}

We analyze simulation \ref{MainF} and then compare the results from all our other simulations to that of \ref{MainF}.
We identify the following 4 stages of evolution in our merger simulations:
\begin{enumerate}
    \item[(1)] The pre-plunge stage ($t - t_\text{plunge} < 0$) is marked by a gradual increase in $L_\text{cool}$ over tens of days. 

    \item[(2)] The ``hot peak'' stage ($0 \lesssim t - t_\text{plunge} \lesssim 5\,\text{days}$), during which both $L_\text{obs}$ and $L_\text{cool}$ rise to a local maximum and the weighted average spectrum peak temperature $\langle T_\text{sp}\rangle$ is the highest among all other local $L_\text{obs}$ maxima.

    \item[(3)] The plateau stage ($5 \lesssim t-t_\text{plunge} \lesssim 20\,\text{days}$), during which the value of $L_\text{obs}$ stays within about an order of magnitude.

    \item[(4)] The dimming stage ($t - t_\text{plunge}\gtrsim 20\,\text{days}$) features a decline in $L_\text{obs}$ and either a global or local maximum in $L_\text{cool}$ tens of days after the plateau stage.
\end{enumerate}

Figure~\ref{fig:Fcellimagesangles} shows a visualization of the observed flux during these four stages and at the moment of the plunge-in, from three different observation angles.
We show the entire simulation domain in Figure~\ref{fig:Fcellimagesangles}.
The corresponding light curves can be seen in Figure~\ref{fig:main_light_curve}, in which each snapshot uses the adaptive resolution technique described in \S\ref{sec:imaging}.
The flux spectral distribution can be seen in Figure~\ref{fig:spectrums}.
The observed luminosity evolution in the $V$-band, $R$-band, and $I$-band is shown in Figure~\ref{fig:MainF_Lv_Lr_Li}.

\begin{figure*}
    \centering
    \includegraphics{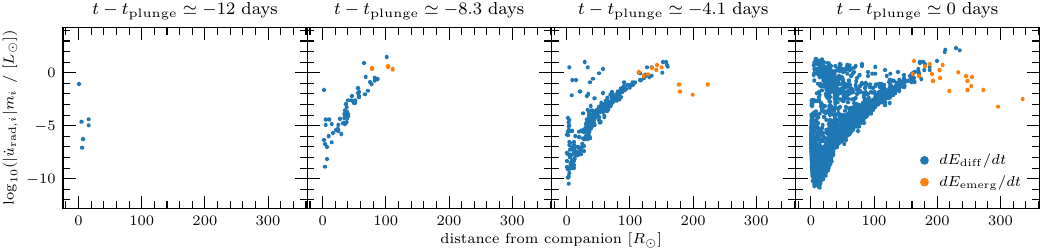}
    \label{fig:MainFdEdtfromsecondary}
    \caption{We show the radiative energy losses of ejected SPH particles during the pre-plunge stage in simulation \ref{MainF} with distance from the companion's center of mass. The center of mass is obtained using only the SPH particles which are bound to the system and which were bound to the companion star prior to Roche lobe overflow. Particles which radiate by $dE_\text{diff}/dt$ are shown in blue, and those which radiate by $dE_\text{emerg}/dt$ in orange.}
\end{figure*}

\subsection{The pre-plunge stage}

During this stage, the donor star gradually loses matter at an increasing rate, which then expands and increases the overall luminosity.
At the start of the simulation, the $\sim 40\,000$ particles that comprise the top $0.1\,M_\odot$ of the donor's envelope each individually have masses about $10^{-6}-10^{-5}\,M_\odot$ (about $2.5\times10^{-6}\,M_\odot$ on average). 
At the end of the pre-plunge stage, $4\,938$ particles have been ejected.
About $\gtrsim 86\,\text{per}\,\text{cent}$ of these ejected particles radiate by $dE_\text{diff}/dt$ and the other $\sim 14\,\text{per}\,\text{cent}$ radiate by $dE_\text{emerg}/dt$, as shown in Figure~\ref{fig:MainFdEdtfromsecondary}.

The total ejected mass during this stage is about $10^{-2}\,M_\odot$.
The particles being ejected can radiate as much as $\sim300\,L_\odot$ each at the peak of their cooling during this stage.
Accordingly, the limited mass resolution leads to an artificial noise in the $L_\text{obs}$ light curve at the start of the rise in $L_\text{obs}$.

Figure~\ref{fig:spectrums} shows the spectral distributions during the evolutionary stages.
As expected, a single BB corresponding to spectrum peak temperature cannot fit the integrated spectrum.
For every stage, the integrated spectral distribution at red wavelengths significantly exceeds that of a blackbody with the same peak temperature.
The most striking difference between the blackbody spectrum and the integrated spectrum is during the pre-plunge stage: if the observer is looking at the system from above ($i=90^\circ$), there is an additional blue peak in the spectrum at $\lambda=160\,\text{nm}$, though it has a lower intensity than the main peak at $\lambda=850\,\text{nm}$.
The blue peak is formed by the transfer of the donor's material into the companion's proximity, i.e. by an interim accretion disk.
This additional blue peak becomes less blue and has a lower intensity than the red peak when viewed from $i=45^\circ$.
During system rotation, the derived (by peak value) spectrum peak temperature can oscillate between red and blue peaks.
As more mass becomes ejected, the two peaks gradually join each other, as shown by the red and cyan curves in Figure~\ref{fig:spectrums}, corresponding to $t-t_\text{plunge}=0\,\text{days}$ and $3\,\text{days}$ respectively.
While the existence of the blue peak affects values of the bolometric luminosity $L_{\rm bol}$, the specific luminosities $L_{\text{obs}, V}$, $L_{\text{obs},R}$, and $L_{\text{obs}, I}$ are almost angle independent, as shown in Figure~\ref{fig:MainF_Lv_Lr_Li}.

\begin{figure}
    \centering
    \includegraphics{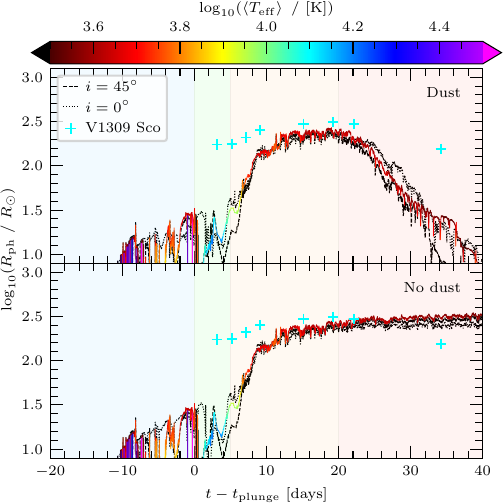}
    \label{fig:MainFrphotosphere}
    \caption{The ``effective" photospheric radius $R_\text{ph}\equiv (L_\text{obs} / 4\pi\sigma \langle T_\text{sp}\rangle^4)^{1/2}$ in \ref{MainF} using dust opacities $\kappa_\text{dust}=1\,\text{cm}^2\,\text{g}^{-1}$ as described in \S\ref{sec:keyquantities} (top panel) and dustless opacities $\kappa_\text{dust}=10^{-4}\,\text{cm}^2\,\text{g}^{-1}$ (bottom panel). We show the V1309~Sco (cyan crosses) photospheric radius $R_\text{ph}$ from table~1 of \citet{2011A&A...528A.114T}, shifted in time so that the maximum value of $L$ in that table corresponds with the maximum $L_\text{obs}$ during the hot peak in \ref{MainF}. The colors of the V1309~Sco data are only aesthetic and do not represent $T_\text{eff}$. The \ref{MainF} evolutionary stages are shown as light blue for pre-plunge, light green for hot peak, light orange for plateau, and light red for dimming.} 
\end{figure}

\begin{figure}
    \centering
    \includegraphics{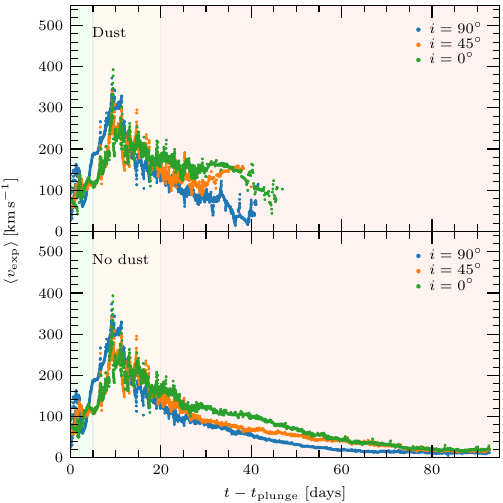}
    \label{fig:MainFvelocities}
    \caption{We show the flux-averaged expansion velocity $\langle v_\text{exp}\rangle$ as defined in Equation~(\ref{eq:averagevexp}) for simulation \ref{MainF}, where the $+\hat{z}$ direction is pointing toward the observer, who is located at inclination angles $i=90^\circ$ (blue), $i=45^\circ$ (orange), and $i=0^\circ$ (green). We use dust opacities $\kappa_\text{dust}=1\,\text{cm}^2\,\text{g}^{-1}$ (top panel) and dustless opacities $\kappa_\text{dust}=10^{-4}\,\text{cm}^2\,\text{g}^{-1}$ (bottom panel). We exclude from the plot any data which corresponds to $L_\text{obs} \leq 1\,L_\odot$. The \ref{MainF} evolutionary stages are shown as light blue for pre-plunge, light green for hot peak, light orange for plateau, and light red for dimming.}
\end{figure}

\subsection{The hot peak stage} \label{sec:hotpeakstage}

The companion enters the envelope during this stage.
Shocks are expected to be produced inside the donor's envelope during the spiral-in and propagate to the surface.
However, most of the flux the observer can detect during this stage comes from the material in the accretion disk nearby the companion and from the asymmetric, so-called ``initial ejecta'', which rapidly leaves the donor and takes away most of the angular momentum \citep{2016MNRAS.462..362I}.
The hot maximum in $L_\text{obs}$ can undergo a dip as either all the initial ejecta has become transparent after passing its maximum in luminosity or the cooling asymmetric ejecta starts to obscure the earlier ejecta.

When the rate at which particles are ejected increases to about $12\,000$ per day, an exponential rise occurs in $L_\text{obs}$ from about $300$ to $1.3\times10^4\,L_\odot$ during $0 < t-t_\text{plunge}\lesssim 3\,\text{days}$, as shown in Figure~\ref{fig:main_light_curve}.
The rise of the light curve ends with a hot peak, which is characterized by a larger spectrum peak temperature than in subsequent $L_\text{obs}$ peaks.

The spectrum peak temperatures during this stage can exceed $9\,000\,\text{K}$.
The secondary peak in spectral distribution that can be seen during the pre-plunge stage (Figure~\ref{fig:spectrums}) increases in relative intensity.
For a brief period of time near the start of this stage, the spectrum peak temperatures can be identified sporadically at about $30\,000\,\text{K}$.
As the two blue and red peaks join, the spectrum peak temperature gradually increases to about $15\,800 \pm 800\,\text{K}$ and then gradually decreases to about $9\,000\,\text{K}$, as shown in Figure~\ref{fig:Teff_MainF_MainFP_MainFR}.
The angular dependence of the light curve and spectrum peak temperature is the smallest during this stage when compared to the other stages.

The flux-weighted average expansion velocity $\langle v_\text{exp}\rangle$ linearly increases as well, from about $50\,\text{km}\,\text{s}^{-1}$ to about $200\,\text{km}\,\text{s}^{-1}$ for $i=90^\circ$ at $t-t_\text{plunge}\approx5\,\text{days}$, as shown in Figure~\ref{fig:MainFvelocities}.
When the flux starts to become dominated by low-temperature, dust-forming ejecta during the dimming stage, the velocities $\langle v_\text{exp}\rangle$ seen by the observer from $i=0^\circ$ (edge on to the orbital plane) are revealed to be about twice those seen from $i=90^\circ$ (top down).

The approximately $8\,200$ particles that simultaneously contribute to $L_\text{obs}$ at $i=0^\circ$ during the hot peak stage are mainly located within $340\,R_\odot -470\,R_\odot$, and their kernels extend no further than $420\,R_\odot-560\,R_\odot$.
The vast majority of these particles radiate diffusely, via $dE_\text{diff}/dt$, for about $97\,\text{per}\,\text{cent}$ of the total radiation from these particles.
We note that the concept of ``effective'' photospheric radius $R_\text{ph}$ during this stage is misleading, as it suggests that the photosphere is located at a smaller distance than the location of emitting particles during the hot peak stage.

\subsection{The plateau stage}\label{sec:theplateaustage}

This evolutionary stage occurs when the material ejected during the plunge-in has adiabatically cooled to the regime where it can radiate its energy away effectively.
After an SPH particle radiates away (the cooling is faster than adiabatic), it becomes transparent, allowing the observer to see the next SPH particle behind it (if there is one).
The ejected material initially expands almost adiabatically until its temperature is low enough for radiative cooling to become the most influential determinant of temperature change.
The time it takes to expand adiabatically explains a slight delay between the hot peak and plateau stages.
Specifically, at about $t-t_\text{plunge}\gtrsim10\,\text{days}$ in \ref{MainF}, the early asymmetric ejecta has passed its maximum luminosity production, leading to a slight dip after the hot peak.
The production of luminosity then switches to being supported by the plunge-in ejecta.

In general, the entropy of the ejecta from a merger is about uniform for each considered merger setup, while it can differ between the simulations.  The value of that entropy dictates how far the ejecta must travel in order for radiative cooling to become more efficient than adiabatic cooling.
Hence, the radiating surface appears at about the same distance from the initial binary.
As a result, after the ejecta has first reached that distance, the luminosity remains about constant for as long as the ejecta is gradually becoming transparent (but see below on the dimming).
We observe the same behavior in \ref{MainF}.

The ``effective'' photosphere, $R_\text{ph} \equiv (L_\text{obs}/4\pi\sigma\langle T_\text{sp}\rangle^4)^{1/2}$, increases exponentially at the start of this stage, as shown in Figure~\ref{fig:MainFrphotosphere}, with no noticeable angular dependence.
The linear (in log) increase completes with a radius of about $160\,R_\odot$.

We reiterate here that $R_\text{ph}$ value is irrelevant to the location where the observed radiation is produced. The shift from optically thick to thin regime is very gradual. On average, the observed flux here is formed by $15$ particles along each ray, counted until the optical depth along the ray is about $40$.
Most of the radiation during the plateau phase comes from particles of $0.05 < \tau_i < 20$. On average, six particles form a $\delta\tau \approx 0.1$ (from $0.01$ to $0.15$) layer along each ray between the most radiating particles and the observer, simultaneously contributing to the overall flux while attenuating the brightness of the particles along the ray.

In more detail, after going through the hot peak, from $t-t_\text{plunge}\approx 7\,\text{days}$, simulation \ref{MainF} gradually increases in $L_\text{obs}$ to about the same as that during the hot peak, but at much cooler spectrum peak temperatures, about $4\,800\pm800\,\text{K}$.
The derived ``effective'' photospheric radius $R_\text{ph}$ slowly increases to about $250\,R_\odot$.
In this stage, the flux-weighted expansion velocities $\langle v_\text{exp} \rangle$ reach a maximum and then enter a gradual decline.
Particles ejected during the early stage of the plunge-in typically have expansion velocities about twice that of particles ejected earlier, although some are slower than the particles ejected earlier.
The initiated plunge-in ejecta overtakes the early ejecta within about $10\,\text{days}$ after the plunge-in, such that $\langle v_\text{exp}\rangle$ increases due to the flux becoming dominated by the faster ejecta.
For example, for an inclination angle $i=0^\circ$, $\langle v_\text{exp} \rangle \approx 200\,\text{km}\,\text{s}^{-1}$ at $20\,\text{days}$ after the plunge, and proportionally smaller $\langle v_\text{exp} \rangle$ are observed from other inclination angles, about $150\,\text{km}\,\text{s}^{-1}$, as shown in Figure~\ref{fig:MainFvelocities}.

After $t-t_\text{plunge}\gtrsim10\,\text{days}$, the observer sees an expanding gas cloud that expands radially outward with time.
The remnant's shape stays approximately constant as it expands.

\subsection{The dimming stage}

During this evolutionary stage, the observed bolometric luminosity $L_\text{obs}$ starts to noticeably decrease compared to the plateau luminosity.
Depending on the adopted physical assumptions, there are two types of dimming: ``dusty'' dimming and ``transparency'' dimming.
Dusty dimming is observed if it is assumed that dust can form; the newly formed dust starts to obscure the merged star and the remaining, still expanding ejecta.
Transparency dimming is when no dust formation is assumed; once all the ejected material radiates away, it cools down, becomes transparent, and reveals the merged star.

When the dimming is dusty, $L_\text{obs}$ decreases strongly and quickly.
We observe dusty dimming for \ref{MainF} only if we assume that dust forms immediately, such that opacities in matter with $100 < T < 1\,000\,\text{K}$ is non-negligible.
The deviation in $L_\text{obs}$ between the cases with and without dust starts at about $25\,\text{days}$ after the plunge, or about $15\,\text{days}$ after the start of the plateau stage.
In \S\ref{sec:roleofopacities} we consider the case in which cold temperature opacities are either Planck or Rosseland mean values.

A result of dusty dimming is a contraction of the ``effective'' photosphere from $\sim220\,R_\odot$ to a few solar radii, as shown in Figure~\ref{fig:MainFrphotosphere}.
The fastest ``effective'' photospheric contraction occurs with $i=45^\circ$.
The spectral distribution of the flux at this viewing angle has the most significant red excess.

With transparency dimming, the spectrum peak temperature drops below $\sim 4\,000\,\text{K}$ at about the same time as dusty dimming.
After further cooling, the temperature remains about the same as that at which the material becomes transparent.
Specifically, at the end of the shown evolution in the ``no dust'' case, the spectrum peak temperature is $\sim 3\,400\,\text{K}$.

In more detail, transparency dimming results in most of the ejecta becoming transparent at about $60\,\text{days}$ after the plunge.
The brightness begins to decrease because the effective ``photosphere'' -- the surface of the still opaque gas -- contracts.
However, the luminosity does not fall below about $1\,000\,L_\odot$ for some time because the donor star is significantly out of thermal equilibrium.
As expected for a merged star, the material that was not ejected extends to about $100\,R_\odot$, with only about $10^{-2}\,M_\odot$ remaining bound outside of $100\,R_\odot$.
This bound material cools slowly compared to the simulation timescale.
As a result, we only achieve complete dimming when dust formation is assumed.
The spectrum peak temperature of this long-term glow stays near $3\,400\,\text{K}$ (flux intensity is at $95\,\text{per}\,\text{cent}$ of the maximum intensity between $2\,850\,\text{K}$ and $3\,970\,\text{K}$), and the corresponding luminosity stays at about $1\,200\,L_\odot$.

The expansion velocities $\langle v_\text{exp}\rangle$ at about $60\,\text{days}$ after the plunge are below $100\,\text{km}\,\text{s}^{-1}$ in the ``no dust'' case.

The evolution of luminosity in the specific bands were almost indistinguishable in the previous stages.
In this stage, we find that the $V$-band luminosity decreases slightly faster; see Figure~\ref{fig:MainF_Lv_Lr_Li}.

\section{Results: role of opacities}\label{sec:roleofopacities}

\begin{figure}
    \centering
    \includegraphics{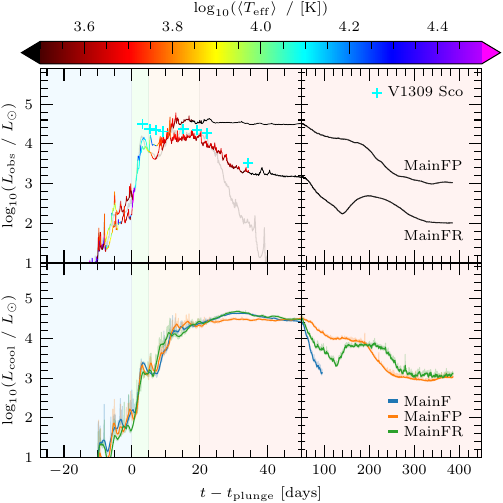}
    \label{fig:MainFPMainFRlightcurves}
    \caption{We show the light curves of observed luminosity $L_\text{obs}$ (top panel) and total radiative losses into empty space $L_\text{cool}$ (bottom panel) for our simulations \ref{MainF}, \ref{MainFP} and \ref{MainFR}. We separate timescales with a vertical axis, where $t-t_\text{plunge}\leq 50\,\text{days}$ is on the left and $t-t_\text{plunge}> 50\,\text{days}$ is on the right. In the top panel \ref{MainF} is shown in gray. In the bottom panel we show \ref{MainF} in blue, \ref{MainFP} in orange, and \ref{MainFR} in green. We show the V1309~Sco (cyan crosses) luminosity $L$ from table~1 of \citet{2011A&A...528A.114T}, shifted in time so that the maximum value of $L$ in that table corresponds with the maximum $L_\text{obs}$ during the hot peak in \ref{MainF}. The colors of the V1309~Sco data are only aesthetic and do not represent $T_\text{sp}$. The \ref{MainF} evolutionary stages are shown as light blue for pre-plunge, light green for hot peak, light orange for plateau, and light red for dimming.}
\end{figure}

\begin{figure*}
    \centering
    \includegraphics{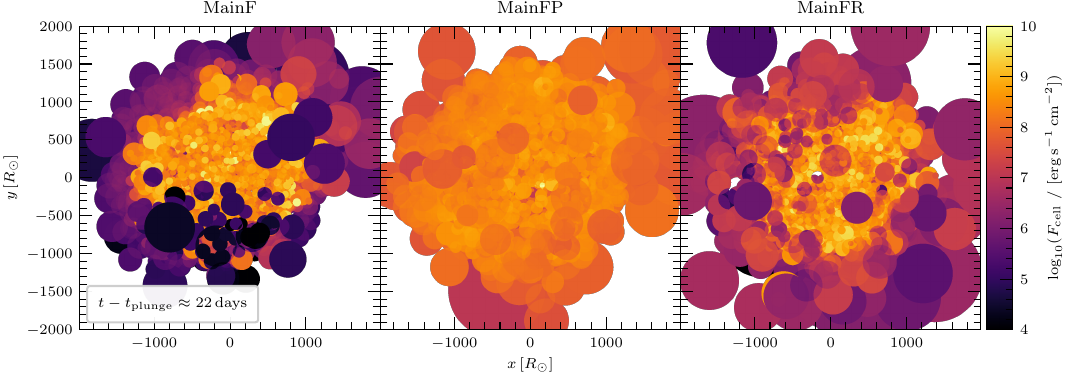}
    \label{fig:flux_image_MainF_MainFP_MainFR}
    \caption{We show the image cell flux $F_\text{cell}$ as described in Equation~(\ref{eq:Fcell}) for simulations \ref{MainF} (left), \ref{MainFP} (middle), and \ref{MainFR} (right) at $t-t_\text{plunge}\approx22\,\text{days}$ in ``screen space'' coordinates $xy$, with the observer located at viewing angle $i=90^\circ$ (``top down'' view on the orbital plane). The images shown here are the same as those used to obtain $L_\text{obs}$ in Figure~\ref{fig:comparisoncoolingcurves}. In \ref{MainF} some of the outermost particles have cooled to $T_i < 1\,000\,\text{K}$, which is where we use our post processing dust opacities $\kappa_\text{dust}=1\,\text{cm}^2\,\text{g}^{-1}$. Dust opacities are not used in \ref{MainFP} and \ref{MainFR}. \ref{MainF} and \ref{MainFR} show a dark halo forming at this time.}
\end{figure*}

\begin{figure}
    \centering
    \includegraphics{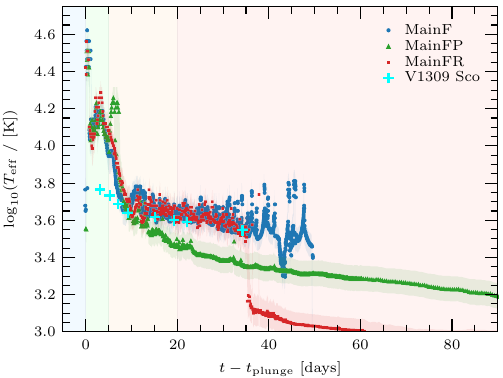}
    \label{fig:Teff_MainF_MainFP_MainFR}
    \caption{We show the spectrum peak temperatures $T_\text{sp}$ for simulations \ref{MainF}, \ref{MainFP} and \ref{MainFR} after the plunge-in. Post processing dust opacities $\kappa_\text{dust}=1\,\text{cm}^2\,\text{g}^{-1}$ are used for \ref{MainF} only. We exclude from the plot data for which $L_\text{obs}<1\,L_\odot$. We show the V1309~Sco (cyan crosses) $T_\text{eff}$ from table~1 of \citet{2011A&A...528A.114T}, shifted in time so that the maximum value of $L$ in that table corresponds with the maximum $L_\text{obs}$ during the hot peak in \ref{MainF}. The \ref{MainF} evolutionary stages are shown as light blue for pre-plunge, light green for hot peak, light orange for plateau, and light red for dimming.}
\end{figure}

As we showed before, the boolean-type assumption of either including or not including dust opacities strongly alters the resulting $L_\text{obs}$ light curves.
During the dynamical evolution, our simulations \ref{MainF}, \ref{MainFP}, and \ref{MainFR} use opacity tables \ref{fullopacities}, \ref{planckonlyopacities}, and \ref{rosselandonlyopacities} respectively, as described in \S\ref{sec:opacities}.
Hereafter, we call the set made of \ref{MainF}, \ref{MainFP}, and \ref{MainFR} as our ``opacity simulations''. 
The opacities in our opacity tables are significantly different for temperatures $\log T \lesssim 3.46$.
The light curves for \ref{MainFP} and \ref{MainFR} can be seen in  Figure~\ref{fig:MainFPMainFRlightcurves}.

The ejecta remnant seen by the observer produces non-uniform radiative flux.
The cells in the images have lower characteristic temperatures $T_\text{cell}$ (see Equation~\ref{eq:fluxweightedaveragecell}) around the remnant in its ``dark halo'': near the edge of the remnant, the image rays first pass through particles which are cooler and more transparent than those farther away along the line of sight (Figure~\ref{fig:flux_image_MainF_MainFP_MainFR}).
The size of the halo and its brightness depend on the adopted opacities.

Specifically, in simulation \ref{MainF}, the opacity of the particles during the dynamical simulation -- not the image post processing -- becomes negligible ($\kappa_i = 10^{-4}\,\text{cm}^2\,\text{g}^{-1}$) as the temperature decreases below $3\,000\,\text{K}$.
As a result, when the particles near the edge of the expanding remnant cool down to the transition temperature (almost adiabatically, despite ongoing radiative losses), they become optically thin.
The transition from optically thick to optically thin is when most of the radiative losses take place, and the particles temporarily cool down noticeably faster than through adiabatic cooling.

Consequently, until the outermost ejecta cools to $3\,000\,\text{K}$, the remnant appears to have an approximately uniform temperature distribution to within about an order of magnitude and the halo is insignificant.
The halo forms at the transition temperature.
Although the halo is observed at the outer edge, it is present in all directions, and it does not contribute much to the image rays that eventually encounter hot and optically thick particles.
Once the outermost particles cool to temperatures below $1\,000\,\text{K}$, their opacities increase to the adopted constant dust opacity, $\kappa_\text{dust} = 1\,\text{cm}^2\,\text{g}^{-1}$.
The entire remnant quickly becomes obscured by dust and $L_\text{obs}$ drops.
This drop in luminosity is purely driven by the dust opacity and can only be present when dust is formed immediately once the temperature falls below the $1\,000\,\text{K}$ threshold.
For each simulation, we apply $\kappa_\text{dust}$ only when creating the post processed $L_\text{obs}$ images; $\kappa_\text{dust}$ is not used during the dynamical evolution.

A similar transition is observed in simulation \ref{MainFR}.
Here, opacities within $1\,200\,\text{K} \lesssim T_i\lesssim 3\,000\,\text{K}$ are even smaller than those in \ref{MainF}, but are non-negligible for $T\lesssim 1\,200\,\text{K}$ during the simulations.
As a result, $L_\text{obs}$ drops sooner in \ref{MainFR} than in \ref{MainF} due to a more rapid decline in the spectrum peak temperature; see Figure~\ref{fig:Teff_MainF_MainFP_MainFR}).

The non-negligible opacities for $T\lesssim 1\,200\,\text{K}$ in \ref{MainFR} gradually slows the cooling back down to the adiabatic regime, and the $L_\text{obs}$ decline becomes gradually less steep.
The temperature of the object during this time appears to be the same as at the low end of the transparency window, see Figure~\ref{fig:Teff_MainF_MainFP_MainFR}.

In contrast, in \ref{MainFP}, opacities in the transparency window drop with decreasing temperature less significantly than in \ref{MainFR}.
The spectrum peak temperature decreases smoothly, without a sharp transition, and dust-permitting temperatures are not achieved within the considered $60\,\text{days}$.
The $L_\text{obs}$ light curve declines very gradually, with a similar gradual decline in spectrum peak temperature.

In \ref{MainF}, $L_\text{obs}$ drops below $1 L_\odot$ about $40\,\text{days}$ after the plunge-in when dust opacities are considered, but \ref{MainFR} and \ref{MainFP} remain bright for hundreds of days after until their spectrum peak temperature is very low.
\ref{MainFR} and \ref{MainFP} can be observed as ``warm dust''.

\section{Results: role of varied physics.}\label{sec:roleofvariedphysics}

\begin{figure*}
    \centering
    \includegraphics{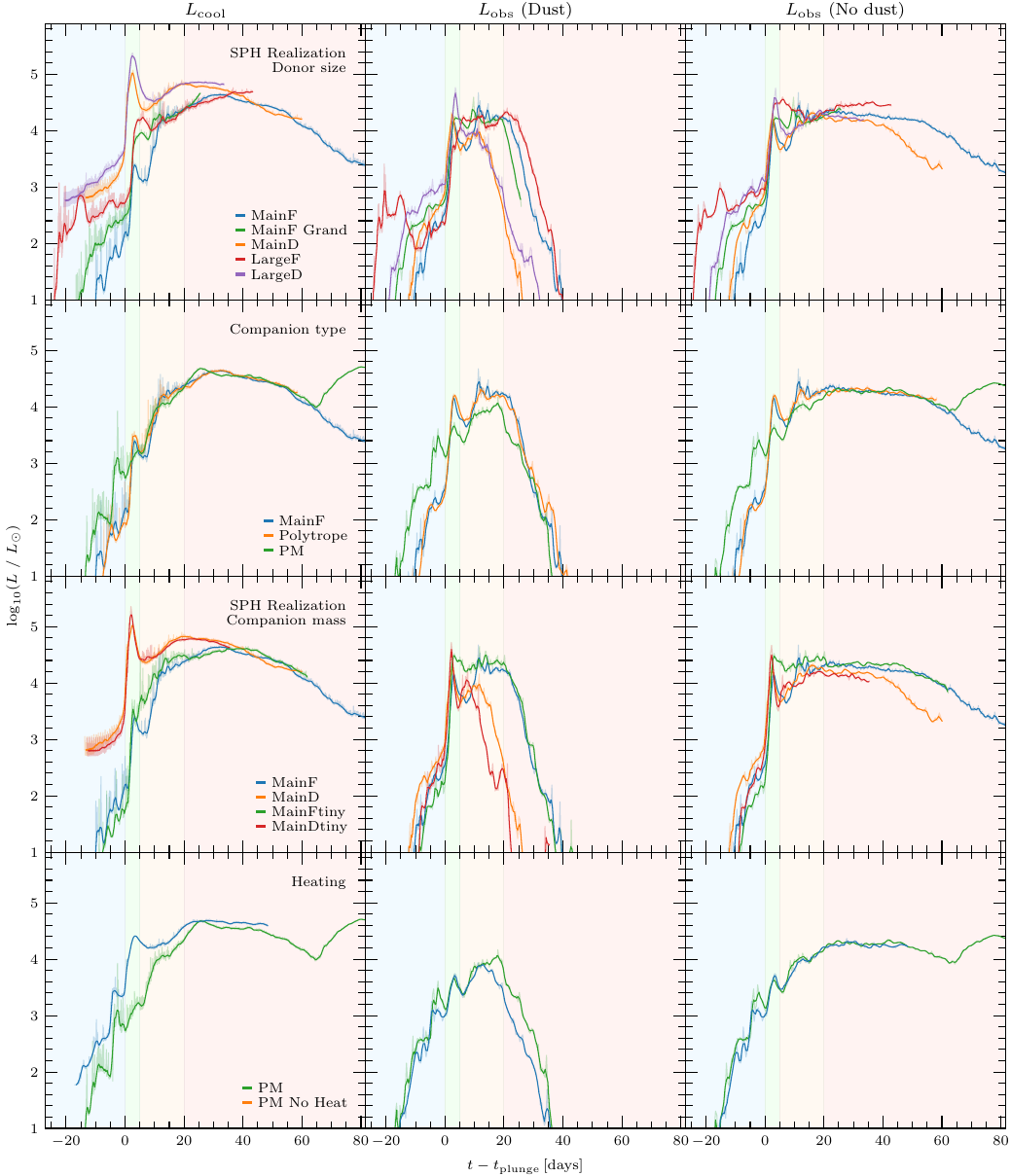}
    \label{fig:comparisoncoolingcurves}
    \caption{For each of our simulation comparisons in Table~\ref{tab:comparisons} we show the total radiative losses into empty space $L_\text{cool}$ (left), the observed luminosity $L_\text{obs}$ using our post-processing dust opacity $\kappa_\text{dust}=1\,\text{cm}^2\,\text{g}^{-1}$ (middle), and no post-processing dust opacity (right). We do not show the opacity comparison in this figure (see Figure~\ref{fig:MainFPMainFRlightcurves}). We show a time-average (opaque lines) with a window of $1\,\text{day}$ at each snapshot (semitransparent lines). That is, for each snapshot we let $L_\text{cool}$ be the average of the $L_\text{cool}$ from each snapshot within a $1\,\text{day}$ window. The time average is taken simultaneously for each snapshot. We also show the time average for $L_\text{cool}$. The \ref{MainF} evolutionary stages are shown as light blue for pre-plunge, light green for hot peak, light orange for plateau, and light red for dimming.}
\end{figure*}

\begin{figure}
    \centering
    \includegraphics{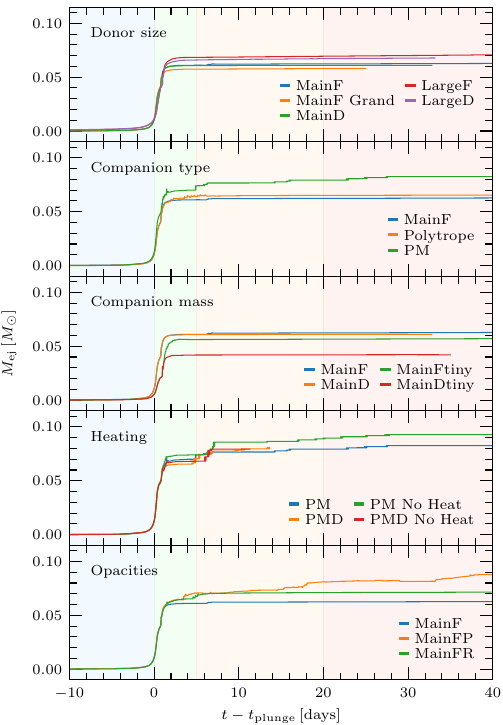}
    \label{fig:mejecta}
    \caption{We show the time evolution of the ejected mass $M_\text{ej}$ in $M_\odot$. Each panel corresponds to one of the comparisons in \S\ref{sec:simulations}. The \ref{MainF} evolutionary stages are shown as light blue for pre-plunge, light green for hot peak, light orange for plateau, and light red for dimming.}
\end{figure}

In this section, we will analyze how different physics can impact the resulting light curves, without varying the opacities.
The various sets of light curves generated from our simulations are depicted in Figure~\ref{fig:comparisoncoolingcurves} and the evolution of the ejected mass in Figure~\ref{fig:mejecta}.

\subsection{Role of type of SPH realization}

Hereafter, we name the following simulations as ``fluffy simulations'': \ref{MainF}, \ref{LargeF}, \ref{MainFtiny}, \ref{PM}, and \ref{PMNoHeat}.
Each fluffy simulation has a corresponding ``dense simulation'': \ref{MainD}, \ref{LargeD}, \ref{MainDtiny}, \ref{PMD}, and \ref{PMDNoHeat}, respectively.

Most simulation pairs considered here have no noticeable difference in the amount of ejected mass in the fluffy simulation compared to their dense simulation counterparts, apart from the case of \ref{MainFtiny} and \ref{MainDtiny}.
The change in plunge-in times can be up to about $6\,\text{days}$, as in \ref{LargeF} and \ref{LargeD}.

Prior to the dimming stage, we observe that $L_\text{cool}$ is larger in our dense simulations than in our fluffy simulations.
The difference is especially noticeable during the hot peak stage, see Figure~\ref{fig:comparisoncoolingcurves}.
The reason is that, for each cooling ray in an SPH particle, the maximum possible cooling rate corresponds with the maximum possible temperature gradient, which is achieved when no other particles are along a ray's path.
This maximum cooling rate is usually obtained by cooling rays that originate from particles near the surface of the gas cloud and point in the outward direction.
Particles in the dense cooling regime are ``smaller'' than those in the fluffy regime, which has two effects: first, rays that emerge from smaller particles have a lower probability of finding other obstructing particles, and second, an unobscured ray has a temperature gradient that is inversely proportional to the particle size.

Our dense simulations decline in $L_\text{obs}$ earlier than in the corresponding fluffy simulations when $L_\text{obs}$ is created with the assumption about immediate dust formation in our flux processing.
This decline is triggered when the outer particles cool below $1\,000\,\text{K}$, where the dust opacity is adopted (see \S\ref{sec:roleofopacities}).
The enhanced radiative cooling efficiency of the surface particles in the dense regime allows dust opacities in the flux processing to be used earlier, leading to a faster decline in $L_\text{obs}$.

When the immediate dust formation assumption is not used, the decline in $L_\text{obs}$ in the dense simulations occurs earlier than in fluffy simulations. For example, the decline from the main plateau takes place at about $30\,\text{days}$ after the plunge-in in the case of \ref{MainD}, while at about $60\,\text{days}$ in \ref{MainF}.
With dense cooling, the ejecta particles ``see'' vacuum and start effectively losing their energy with radiation at slightly shorter distances from the merged binary, leading to an earlier decline.

\subsection{Role of companion mass}\label{sec:roleofcompanionmass}

The $0.06\,M_\odot$ decrease in the companion mass between simulations \ref{MainF} and \ref{MainD} and \ref{MainFtiny} and \ref{MainDtiny} results in a sooner plunge-in by about $18\,\text{hours}$.
\ref{MainFtiny} ejects about $10\,\text{per}\,\text{cent}$ less mass than \ref{MainF}, and \ref{MainDtiny} ejects about $30\,\text{per}\,\text{cent}$ less mass than \ref{MainD}.

The $L_\text{cool}$ light curves evolve very similarly in the case of a smaller mass companion.
A slight discrepancy can be seen in the early plateau stage in \ref{MainF} and \ref{MainFtiny}, $5\lesssim t-t_\text{plunge}\lesssim 10\,\text{days}$.
Here, the smaller companion surprisingly produces both more cooling and more observed luminosity.

The ejecta velocities are about the same between the two cases, with slightly smaller velocities overall in the simulations that have have the less massive companion.
However, if we apply a weighting to the velocities of the ejecta particles where the weights are their cooling rates, we find \ref{MainFtiny} and \ref{MainDtiny} have larger weighted velocities than their more massive companion counterparts.
This reverse velocity behavior seems to be counterintuitive.
We note that the velocities weighted by the cooling rate are formed by particles that were adiabatically cooled to the temperatures where radiative cooling becomes efficient.
The ejected particles in the simulations with a less massive companion have less shock heating and smaller entropies.
Hence, the particles in \ref{MainFtiny} start to radiate away earlier, and at smaller distances to the binary, at the location where their velocities are still larger than in \ref{MainF}.

\subsection{Role of companion type}\label{sec:roleofcompaniontype}

We find negligible differences in $t_\text{plunge}$, $M_\text{ej}$, $L_\text{cool}$, and $L_\text{obs}$ between the simulation with the polytrope companion, \ref{Polytrope}, and the simulation with the \texttt{MESA} companion, \ref{MainF}, as shown in Table~\ref{tab:comparisons} and Figure~\ref{fig:comparisoncoolingcurves}.

However, our point mass companion simulation \ref{PM} presents notable differences compared to \ref{MainF} and \ref{Polytrope}.
In \ref{PM}, $t_\text{plunge}$ is prolonged by several days, and approximately $30\,\text{per}\,\text{cent}$ more mass is ejected.
The prolonging of the plunge-in occurs because tidal forces with a point mass are less efficient.
The increase in ejected mass occurs because particles experiencing close interactions with a point mass attain higher velocities than they would with a non-compact cloud of equivalent total mass, and hence more particles can become unbound.

We find that at $10\,\text{days}$ after the plunge the average ejecta velocity is slightly higher and average ejecta entropy is slightly lower in \ref{PM} than in \ref{MainF} and \ref{Polytrope}.
At the same time, the average specific internal energy of the ejecta is higher in \ref{MainF} and \ref{Polytrope} than in \ref{PM}, indicating that there was less shock heating in \ref{PM}.
With higher velocities and smaller entropies, the dust is formed sooner in \ref{PM} than in \ref{MainF} and \ref{Polytrope}, and the dimming starts earlier (see Figure~\ref{fig:comparisoncoolingcurves}).

In the ``no dust'' case, $L_\text{obs}$ is the same in \ref{PM} and \ref{MainF} during the dimming stage until $t-t_\text{plunge}\gtrsim60\,\text{days}$.
At this time, about $0.06\,M_\odot$ becomes ejected in \ref{PM}.
The particles that comprise this new ejecta are located deep within the remnant's envelope, have high temperatures, and quickly reach the surface where they can more effectively radiate, directly affecting both $L_\text{obs}$ and $L_\text{cool}$.
We conjecture that this new, dredged-up ejecta is a result of close interactions with the donor and companion core particles, which at this time have a separation of about $0.06\,R_\odot$.
We find that total energy remains conserved in \ref{PM} to within about $99\,\text{per}\,\text{cent}$ during this event.

\subsection{Role of heating}\label{sec:roleofheating}

Our radiative heating prescription as defined in \S\ref{sec:heating} is evaluated using a point mass companion in simulations \ref{PM}, \ref{PMD}, \ref{PMNoHeat}, and \ref{PMDNoHeat}, which we hereafter refer to as our ``heating simulations''.

If we compare the simulations with and without heating, then $L_\text{cool}$ and $L_\text{obs}$ are about the same during the plateau and dimming stages, see Figure~\ref{fig:comparisoncoolingcurves}.
The differences in the light curve and cooling luminosity evolution can be noticed during the plunge-in stage, where simulations with no heat show noticeably stronger cooling, indicating that re-heating is most intense during this stage.
The pre-plunge behavior is less systematic and can be the subject of resolution.
During the plateau stage, simulations with and without heating have about the same properties, indicating that re-heating is not very important.
The cooling luminosity remains about the same.
However, the particles which were ejected right after the plunge become re-heated in our heating simulations, which postpones dust formation and leads to a slightly delayed $L_\text{obs}$ decline.
This is not observed if dust formation is not adopted.

\subsection{Role of resolution}

\begin{figure}
    \centering
    \includegraphics{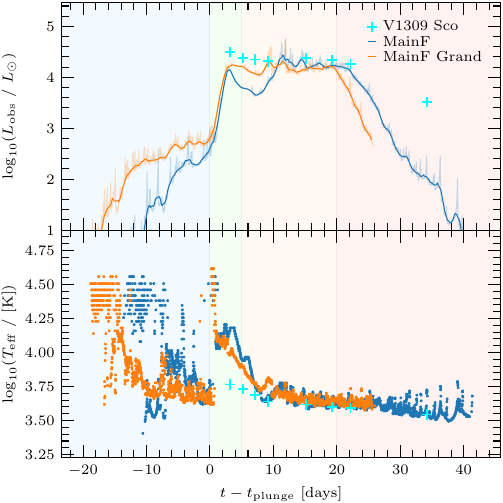}
    \caption{We show the $L_\text{obs}$ light curve (top) and spectrum peak temperature $T_\text{sp}$ evolution (bottom) for our simulations \ref{MainF} (blue) and \ref{MainFGrand} (orange) with dust opacities $\kappa_\text{dust}=1\,\text{cm}^2\,\text{g}^{-1}$ used in the post processed images. We omit data from both panels where $L_\text{obs}<1\,L_\odot$. The time-average (opaque lines) is done in the same way as in Figure~\ref{fig:comparisoncoolingcurves}, with each snapshot shown (semitransparent lines). We show the V1309~Sco (cyan crosses) $L$ and $T_\text{eff}$ from table~1 of \citet{2011A&A...528A.114T}, shifted in time so that the maximum value of $L$ in that table corresponds with the maximum $L_\text{obs}$ during the hot peak in \ref{MainF}. The \ref{MainF} evolutionary stages are shown as light blue for pre-plunge, light green for hot peak, light orange for plateau, and light red for dimming.}
    \label{fig:MainF_MainFGrand_lightcuve_spectra}
\end{figure}

We compare light curves generated for \ref{MainF} and for \ref{MainFGrand}, see Figure~\ref{fig:MainF_MainFGrand_lightcuve_spectra}.
Both their light curves exhibit fluctuations of similar amplitude and frequency during the pre-plunge stage.

A significant fraction of the flux in \ref{MainF} at the $L_\text{obs}$ maximum during the hot peak stage originates from a small collection of particles in the accretion disk, as shown in Figure~\ref{fig:Fcellimagesangles}.
These particles become hidden behind the ejecta during $3\lesssim t-t_\text{plunge}\lesssim 8\,\text{days}$.
However, in \ref{MainFGrand} there is a similar collection of hot particles in the accretion disk, but this hot cluster does not become obscured as it does in \ref{MainF}.
The result is an approximately constant $L_\text{obs}$ after the initial peak.

At the onset of the dimming stage, \ref{MainFGrand} drops in $L_\text{obs}$ sooner than \ref{MainF}.
At $t-t_\text{plunge}\approx20\,\text{days}$ the surface particles begin to form dust in both simulations.
Although these dust-forming particles have larger sizes than those in \ref{MainFGrand}, there are more of such particles in \ref{MainFGrand} than in \ref{MainF}.
The result is that the surface particles in \ref{MainFGrand} are able to more effectively obscure the underlying flux from the observer's view during the dimming stage.

\section{Comparison with V1309~Sco}\label{sec:comparisonwithv1309sco}

The observations of V1309~Sco show a luminosity plateau after initial rise at about $10^{4.4}\,L_\odot$ for about $20\,\text{days}$ observed by AAVSO.
Another few observations by OGLE/SAAO about $770\,\text{days}$ thereafter showed that the luminosity of declined to several $L_\odot$ \citep{2011A&A...528A.114T}.
We find nearly identical plateau features in our observed luminosity $L_\text{obs}$ light curves for simulations \ref{MainF}, \ref{MainFR}, \ref{LargeF}, \ref{MainFtiny}, \ref{Polytrope}, and \ref{MainFGrand}.
During the plateau and dimming stages in \ref{MainF}, the evolution of spectrum peak temperature and photospheric radius is in good agreement with the V1309~Sco observations; see Figures~\ref{fig:MainFrphotosphere} and \ref{fig:Teff_MainF_MainFP_MainFR}. 

When the companion is a compact object, we are unable to produce a synthetic light curve that matches the V1309~Sco luminosity peak.
Each of our point-mass companion simulations \ref{PM}, \ref{PMD}, \ref{PMNoHeat}, and \ref{PMDNoHeat} have light curves which contain a gradual rise to maximum rather than the rapid rise to maximum seen in our other simulations.
Likewise, each have $L_\text{obs}\lesssim 10^4\,L_\odot$, which is dimmer than the V1309~Sco observations by about $2\times10^4\,L_\odot$.

It was argued that dust in the V1309~Sco progenitor was destroyed during the merger, and that new dust was formed at the beginning of the fast optical decline \citep{2016A&A...592A.134T}.
In \ref{MainF} both before and during the merger, the flux peaks in the ultraviolet, as shown in Figure~\ref{fig:spectrums}, which could encourage dust destruction, though we do not include pre-formed dust in our progenitor models.
When we implement our aggressive dust formation regime in our post processed images (see \S\ref{sec:roleofopacities}), we observe that dust begins to form at the onset of the dimming stage.

Without our rapid dust formation regime, our light curves show significantly extended plateaus and never decrease as strongly as that observed in V1309~Sco.
In \ref{MainF}, the luminosity that remains after the decline of the plateau is about $1\,000\,L_\odot$. 
This luminosity is produced by the merged star, which is out of thermal equilibrium and will remain luminous for a long time.

V1309~Sco formed an estimated $5.2\times10^{-4}\,M_\odot$ of dust \citep{2013ApJ...777...23Z} with a temperature of about $800\,\text{K}$ \citep{2013MNRAS.431L..33N}.
We measure the dust mass in our simulations as $\sum_i m_i Z_i$, where each particle $i$ that contributes to the sum has $100<T_i<1\,000\,\text{K}$ and $Z_i=0.02$.
At the end of time integration in each of our simulations we find that the dust mass is within about a factor of two that of V1309~Sco.
In \ref{MainFP} and \ref{MainFR} the dust cools over the span of about a year and thereafter remains approximately constant in temperature at about $650\,\text{K}$ and $540\,\text{K}$ respectively.
Our other simulations did not evolve long enough for the dust temperatures to become constant, though each cooled to temperatures less than $800\,\text{K}$ about $20\,\text{days}$ after the plunge-in.

\section{Conclusions}\label{sec:conclusions}

We present the numerical implementation of FLED radiative transport within an SPH method, which is accomplished by modifying the energy equation.
In the considered binary merger, the impact of radiation on the momentum equation is negligible, with the radiation-induced acceleration being, on average, only $10^{-4}$ of the local gravitational acceleration. Although insignificant in this case, this term could become relevant for other stellar mergers or common envelope events, and can be considered in future applications.

Our investigation involved two distinct approaches for handling particle radiation: the ``fluffy'' method, where particles can overlap in space, generating a shared radiation field, and the ``dense'' method, treating particles as isolated spheres with sizes set by the fluid density.
We observed that the choice of method for calculating the radiative cooling rate had significant impact on the cooling luminosity during the plunge-in, and less impact thereafter.
The particles near the surface of the simulated fluid are disproportionately affected, as the dense method allows these particles to cool more rapidly than they would in the fluffy approach.
The dense cooling approach might yield higher cooling rates than in reality, see \S\ref{sec:dense} for details.
Surprisingly, in most cases, the total amount of ejected mass was only negligibly affected.

We introduced the fluffy approach as the most realistic model, while the dense approach was implemented as a limiting case to evaluate the effect of the maximum plausible cooling rate. Surprisingly, the plateau values of luminosity remain relatively stable regardless of the chosen approach. However, the duration of the plateau is influenced by the rate of dust formation, with shorter durations observed under the dense approach due to faster cooling. Since the dense approach serves as a limiting case, we plan to use only the fluffy approach in future work.

We introduce a method for extracting the luminosity which a real observer would measure.
This involves determining the flux from the object in a specific direction by integrating inward until the contribution of SPH particles in the flux becomes negligible.
By aggregating the spectral contribution of each SPH particle in the image, we can derive the spectral distribution produced by the object.
This obtained spectral distribution enables us to derive not only bolometric luminosities but also luminosities in the $V$-band, $R$-band, and $I$-band, ascertain the spectrum peak temperature of the object, and obtain an ``effective'' photospheric radius for the object.

We applied our method to generate light curves of stellar mergers for progenitor binary systems that were proposed to resemble V1309~Sco.
Across all cases, our simulations consistently traverse the same key stages of evolution, as reflected in the resulting light curves: the plunge-in stage, characterized by an exponential rise in bolometric luminosity; the hot peak stage, where the derived spectrum peak temperature ranges between $9\,000$ and $16\,000\,\text{K}$; the plateau stage, during which the bolometric luminosity remains relatively stable while the spectrum peak temperature is approximately $5\,000\,\text{K}$ gradually decreasing to $4\,000\,\text{K}$; and the dimming stage, marked by a sudden decline in bolometric luminosity which is contingent upon dust opacities and choice of fluffy or dense cooling.
Our findings suggest that some degree of dust formation is likely necessary to account for the rapid luminosity decline observed in LRNe.

In our analysis of the spectral distribution at various evolutionary stages, we compare the integrated spectrum to that of a blackbody with the same temperature as the peak in the integrated spectral distribution.
We observe a consistent red excess in the integrated spectrum.
This red excess phenomenon may potentially result in an overestimation of the bolometric luminosity in observed LRNe.
Prior to the hot peak in the light curve, we also identify the presence of two distinct peaks in the spectral distribution, with one appearing in the red portion and the other in the blue portion.
The presence of ultraviolet radiation prior to the plunge-in may contribute to the destruction of circumstellar dust.

We observe that the expansion velocities can range from approximately $300\,\text{km}\,\text{s}^{-1}$ at a viewing angle of $i=0^\circ$ to about $100\,\text{km}\,\text{s}^{-1}$ at $i=90^\circ$, depending on the phase of the outburst. 
These velocities are expected to gradually decrease if dust opacities do not begin to influence the system.
However, if dust forms rapidly, the far-red spectrum may indicate that the expansion velocities remain consistent with the onset of the plateau stage, as they retain the information about the early ejecta.
We note that the flux which carries that information is very small and might not be detectable.
The temporal evolution in expansion velocities observed in LRNe thus serves as a potential indicator of dust formation timescales.

Through our analysis of our simulations employing various low-temperature opacity models, we ascertain that the decline in our observed light curves during the dimming stage is markedly influenced by opacity.
The opacity effectively dictates the cooling rate of the gas, thereby impacting the luminosity drop observed during this stage.
Additionally, the temperature threshold at which dust formation initiates defines the onset of a second, low-temperature semi-plateau stage.
We find that the temporal evolution of V1309~Sco outburst is best reproduced using the Rosseland opacity table, suggesting the presence of formed dust.

Notably, simulations without dust formation do not exhibit a sharp drop in bolometric luminosity.
The duration of the plateau is prolonged significantly and continues until all unbound particles have radiated away and become transparent.
The merged star is then revealed in our simulations and produces significant flux, which is in contrast to the predicted sharp luminosity drop due to ejecta becoming transparent \citep{2013Sci...339..433I}.
As many stellar evolutionary simulations predicted, the merged star is out of thermal equilibrium and will be luminous for a long time.
The presence of a luminous merger product effectively requires almost immediate dust formation, at least in the case of V1309~Sco.

This result -- the necessity that dust formation be used to explain the observed rapid and significant drop in luminosity -- is in stark contrast to studies which found that dynamical dust formation can take a long time in common envelope ejection \citep{2024arXiv240103644B}.
On the other hand, in simulations that used initially significantly hotter donors, as in \cite{2024ApJ...963L..35C}, dust does not have an effect on the light curve, likely as a result of delayed dust development due to hotter ejecta.
We note that in their simulations, the light curve was continued only down to $L=10^{38}\,\text{erg}\,\text{s}^{-1}$.
This threshold is above the luminosity of the merged products, which were not modelled but were likely thermally disturbed.
If no dust is formed in the case of massive progenitors, we may expect that the light curve would behave as in our ``no dust'' transparency case, with a prolonged plateau, gradual decline, and secondary plateau corresponding to the merged star's luminosity.

If dust formation in the merger ejecta is the most important determining factor to the length of the light curve plateau, then capturing the physics of dust formation may prove crucial for obtaining light curves which match LRNe observations.
Particularly, the formation of ice and molecules may significantly impact the local opacities.
We are hopeful that recent advancements in dust simulation can better capture the fast optical decline observed in LRNe \citep{2024MNRAS.529.4455E}. 

Locating the initial parameters which can reproduce LRNe light curves, such as V1309~Sco in our case, requires both significant computational resources and data management efforts.
For example, the current work required about $300$ CPU years and $50$ GPU years, and produced about $20$ terabytes of data cumulatively, the processing of which required a new, specialized package named \texttt{starsmashertools}\footnote{Available at \href{https://github.com/hatfullr/starsmashertools}{github.com/hatfullr/starsmashertools}.}.
We are hopeful that this suite and others like it will expedite future research.

As more LRNe progenitors are identified \citep{2022MNRAS.517.1884A}, more case-specific simulations can be performed, providing further insights into the physics of stellar mergers.

\vskip1cm

N.I. acknowledges funding from NSERC under Discovery Grant No. RGPIN-2019-04277. 
The authors thank Nadejda Blagorodnova, Tomasz Kaminski, Paul Ricker, Jonathan Dursi,  Jamie Lombardi, Chris Fryer, and Robert Fisher for useful discussions.
The authors thank the anonymous referee for their valuable comments, which have significantly enhanced the quality of the manuscript.
This research was enabled in part by support provided by Prairies DRI and the Digital Research Alliance of Canada (\href{https://alliancecan.ca}{alliancecan.ca}).

\software{StarSmasher \citep{2010MNRAS.402..105G, 2018ascl.soft05010G},
MESA \citep{2011ApJS..192....3P,2013ApJS..208....4P, 2015ApJS..220...15P, 2018ApJS..234...34P, 2019ApJS..243...10P, 2023ApJS..265...15J}}
\facilities{Digital Research Alliance of Canada}

\bibliographystyle{aasjournal}\bibliography{references}

\end{document}